\documentclass[11pt]{article}
\usepackage{epsfig}
\input amssym.def
\input amssym.tex

\newcommand{\qed}{\rule{3mm}{3mm}}

\newcommand{\ed}{{\bf 1}}

\newcommand{\be}{{\bf e}}
\newcommand{\bm}{{\bf m}}
\newcommand{\bL}{{\bf L}}

\newcommand{\bbL}{{\Bbb L}}

\newcommand{\wg}{\widehat{g}}

\newcommand{\cL}{{\cal L}}

\newcommand{\cX}{{\cal X}}

\newcommand{\gog}{{\goth g}}

\newtheorem{theorem}{Theorem}[section]
\newtheorem{proposition}[theorem]{Proposition}
\newtheorem{lemma}[theorem]{Lemma}
\newtheorem{definition}[theorem]{Definition}

\begin{document}

\title{Discrete time Lagrangian mechanics on Lie groups, \\
with an application to the Lagrange top}
\author{A.I.\,Bobenko\footnote{E--mail: {\tt bobenko@math.tu-berlin.de}}
\and Yu.B.\,Suris\footnote{E--mail: {\tt suris@sfb288.math.tu-berlin.de}}}
\date{Fachbereich Mathematik,
Technische Universit\"at Berlin, Str. 17 Juni 136, 10623 Berlin,
Germany}
\maketitle

\begin{abstract}
We develop the theory of discrete time Lagrangian mechanics on Lie groups, 
originated in the work of Veselov and Moser, and the theory of Lagrangian 
reduction in the discrete time setting. The results thus obtained are applied 
to the investigation of an integrable time discretization of a famous integrable
system of classical mechanics, -- the Lagrange top. We recall the derivation
of the Euler--Poinsot equations of motion both in the frame moving with
the body and in the rest frame (the latter ones being less widely known).
We find a discrete time Lagrange function turning into the known continuous
time Lagrangian in the continuous limit, and elaborate both descriptions of
the resulting discrete time system, namely in the body frame and in the rest 
frame. This system naturally inherits Poisson properties of the continuous
time system, the integrals of motion being deformed. The discrete time Lax 
representations are also found. Kirchhoff's kinetic analogy between 
elastic curves and motions of the Lagrange top is also generalised to 
the discrete context.
 
\end{abstract}

\newpage

\section{Introduction}

This paper is devoted to the time discretization of a famous integrable
system of classical mechanics -- the Lagrange top. This is a special 
case of rotation of a rigid body around a fixed point in a homogenious
gravitational field, characterized by the following conditions: the rigid body 
is rotationally symmetric, i.e. two of its three principal moments of inertia 
coincide, and the fixed point lies on the axis of rotational symmetry. 
We present a discretization preserving the integrability property, and
discuss its rich mechanical and geometrical structure. Notice that until
recently \cite{B} only the integrable Euler case of the rigid body motion was
discretized preserving integrability \cite{V}, \cite{MV}, \cite{BLS}.
Consult also \cite{AL}, \cite{H}, \cite{DJM}, \cite{QNCV} for some 
fundamental early papers on the subject of integrable discretizations,
and \cite{BP}, \cite{S} for reviews of this topic reflecting the viewpoints
of the present authors and containing extensive bibliography.

We found the standard presentation of the Lagrange top in mechanical textbooks
insufficient in several respects, and therefore chose to write this paper
in a pedagogical manner, giving a systematic account of the new results
along with the well known ones represented in a form suitable for our present
purposes. The paper is organized as follows.

The introduction recalls the classical Euler--Poinsot 
equations for the motion of the spinning top in the frame moving with the body.
Further, we give less known Euler--Poinsot equations describing the Lagrange
top in the rest frame (they cannot be directly generalized to a general top
case). We finish the introduction by announcing a beautiful time 
discretization of the latter equations. 

In order to derive this discretization
systematically, we need some formalism of discrete time Lagrangian 
mechanics on Lie groups. The discrete time Lagrangian mechanics were introduced
by Veselov \cite{V}, \cite{MV}, see also \cite{WM}, but the case of Lie groups
have certain specific features which, in our opinion, were not worked out
sufficiently. In particular, there lacks a systematic account of the discrete
time version of the Lagrangian reduction (which is fairly well understood
in the continuous time setting, cf. \cite{MS}, \cite{HMR}). Also, we think that 
some technical details in \cite{V}, \cite{MV}, \cite{WM} could be amended: 
in working with variational equations these authors systematically use Lagrange 
multipliers instead of introducing proper notions such as Lie derivatives 
(specific for Lie groups as opposed to general manifolds). Therefore we 
give a detailed exposition of the discrete time Lagrangian mechanics on Lie 
groups in Sect. 3. In order to underline an absolute parallelism of its 
structure with that of the continuous time Lagrangian mechanics, we included 
in Sect. 2 also a presentation of (a fragment of) the latter, which is, of 
course, by no means original.

Sect. 4 is devoted to a Lagrangian derivation of equations of motion of the
Lagrange top, both in the rest and in the body frames. Finally, in Sect. 5
we do the same work for a discrete time Lagrange top. 

It has to be mentioned that the actual motivation for the present development
came from differential geometry, more precisely, from the theory of
elastic curves. A brief account of the relations between spinning tops
and elastic curves is given in Sect. 6.

We also give three appendices. Appendix A is for fixing the notations of 
Lie group theory. In the Appendix B we collect the main results of Sect. 2, 3 
in the form of an easy--to--use table.  Finally, Appendix C contains some 
conventions and simple technical results on a specific Lie group we work with, 
namely on $SU(2)$. It should be remarked here that our experience with various 
integrable discretizations convinced us that working with this group has many 
advantages when compared to the group $SO(3)$, more traditional in this 
context.

A standard form of equations of motion describing rotation of a rigid body 
around a fixed point in a homogeneous gravity field is the following:
\begin{equation}\label{LT body}
\left\{\begin{array}{l}
\dot{M}=M\times \Omega(M)+P\times A\\ \\ \dot{P}=P\times\Omega(M)\end{array}
\right.
\end{equation}
Here $M=(M_1,M_2,M_3)^T\in{\Bbb R}^3$ is the vector of kinetic momentum of the
body, expressed in the so called body frame. This frame is firmly attached to
the body, its origin is in the body's fixed point, and its axes coincide with
the principal inertia axes of the body. The inertia tensor of the body in this
frame is diagonal:
\begin{equation}\label{Inertia}
J=\left(\begin{array}{ccc} J_1 & 0 & 0 \\ 0 & J_2 & 0 \\ 0 & 0 & J_3
\end{array}\right)\;.
\end{equation}
For the vector $\Omega=\Omega(M)$ of the angular velocity we have:
\begin{equation}\label{Omega body}
\Omega=\Omega(M)=J^{-1}M=(J_1^{-1}M_1,J_2^{-1}M_2,J_3^{-1}M_3)^T\in{\Bbb R}^3\;.
\end{equation}
The vector $P=(P_1,P_2,P_3)^T\in{\Bbb R}^3$ is the unit vector along the 
gravity field, with respect to the body frame. Finally, $A=(A_1,A_2,A_3)^T
\in{\Bbb R}^3$ is the vector pointing from the fixed point to the center of 
mass of the body. It is a constant vector in the body frame.

It is well known that the system (\ref{LT body}) is Hamiltonian with respect
to the Lie--Poisson bracket of the Lie algebra $e(3)$ of the Lie group $E(3)$
of euclidean motions of ${\Bbb R}^3$, i.e. with respect to the Poisson bracket
\begin{equation}\label{e3 br body}
\{M_i,M_j\}=\varepsilon_{ijk}M_k\;,\qquad 
\{M_i,P_j\}=\varepsilon_{ijk}P_k\;,\qquad 
\{P_i,P_j\}=0\;,
\end{equation}
where $\varepsilon_{ijk}$ is the sign of the permutation $(ijk)$ of $(123)$.
The Hamiltonian equations of motion for an arbitrary Hamilton function
$H=H(M,P)$ in the bracket (\ref{e3 br body}) read:
\begin{equation}\label{e3 Ham body}
\left\{\begin{array}{l}
\dot{M}=M\times\nabla_M H+P\times\nabla_P H\\ \\ 
\dot{P}=P\times\nabla_M H\end{array}\right.
\end{equation}
which coincides with (\ref{LT body}) if
\begin{equation}\label{LT Ham body}
H(M,P)=\frac{1}{2}\langle M,\Omega(M)\rangle +\langle P,A\rangle\;.
\end{equation}
(Here $\langle\cdot,\cdot\rangle$ stands for the standard euclidean scalar
product in ${\Bbb R}^3$). The Poisson bracket (\ref{e3 br body}) has two Casimir 
functions,
\begin{equation}\label{e3 Cas body}
C=\langle M,P\rangle \qquad {\rm and}\qquad \langle P,P\rangle\;,
\end{equation}
which are therefore integrals of motion for (\ref{LT body}) in involution with
$H(M,P)$ (and with any other function on the phase space). 

The Lagrange case of the rigid body motion (the Lagrange top, for brevity), is
characterized by the following data: $J_1=J_2$, which means that the body is 
rotationally symmetric with respect to the third coordinate axis), and
$A_1=A_2=0$, which means that the fixed point lies on the symmetry axis.
Choosing units properly, we may assume that
\begin{equation}\label{A body}
J_1=J_2=1\;,\qquad J_3=\alpha\;, \qquad A=(0,0,1)^T\;.
\end{equation}
The system (\ref{LT body}) has in this case an additional integral of motion,
\begin{equation}\label{M3 body}
M_3=\langle M,A \rangle\;,
\end{equation}
which is also in involution with $H(M,P)$, and assures therefore the complete
integrability of the flow (\ref{LT body}). For an actual integration of this
flow in terms of elliptic functions see, e.g., \cite{G}, \cite{KS}, and for 
a more modern account \cite{RM}, \cite{Au}, \cite{CB}.

Remarkable as it is, this result is, however, somewhat unsatisfying from the 
practical point of view. Indeed, one is usually interested in describing the
motion of the top in the rest frame, which does not move in the physical space.
It is less known that for the Lagrange top the corresponding equations of motion 
are also very nice and, actually, even somewhat simpler than (\ref{LT body}):
\begin{equation}\label{LT rest}
\left\{\begin{array}{l}
\dot{m}=p\times a\\ \\ \dot{a}=m\times a\end{array}
\right.
\end{equation}
Here $m=(m_1,m_2,m_3)^T\in {\Bbb R}^3$ is the vector of kinetic momentum of the
body, expressed in the rest frame, $p$ is the unit vector along the gravity
field, also expressed in the rest frame, so that it becomes constant:
\begin{equation}\label{p rest}
p=(0,0,1)^T\;,
\end{equation}
and $a=(a_1,a_2,a_3)^T\in{\Bbb R}^3$ is the vector pointing from the fixed point
to the center of mass, expressed in the rest frame. An exterior observer is
mainly interested in the motion of the symmetry axis of the top, which is 
described by the vector $a$. 

The system (\ref{LT rest}) has several remarkable features. First of all, it does
not depend explicitly on the anisotropy parameter $\alpha$ of the inertia tensor.
Second, it is Hamiltonian with respect to the Lie--Poisson bracket of $e(3)$:
\begin{equation}\label{e3 br rest}
\{m_i,m_j\}=-\varepsilon_{ijk}m_k\;,\qquad 
\{m_i,a_j\}=-\varepsilon_{ijk}a_k\;,\qquad 
\{a_i,a_j\}=0\;.
\end{equation}
For an arbitrary Hamilton function $H(m,a)$, the Hamiltonian equations of 
motion in this bracket read:
\begin{equation}\label{e3 Ham rest}
\left\{\begin{array}{l}
\dot{m}=\nabla_m H\times m+\nabla_a H\times a\\ \\ 
\dot{a}=\nabla_m H\times a\end{array}\right.
\end{equation}
These equations coincide with (\ref{LT rest}), if
\begin{equation}\label{LT Ham rest}
H_0(m,a)=\frac{1}{2}\langle m,m\rangle +\langle p,a\rangle\;.
\end{equation}
Of course, the functions
\begin{equation}\label{e3 Cas rest}
c=\langle m,a\rangle \qquad {\rm and}\qquad \langle a,a\rangle\;,
\end{equation}
are Casimirs of the bracket (\ref{e3 br rest}), and therefore are integrals of 
motion for (\ref{LT rest}) in involution with $H_0(m,a)$ (and with any other 
function on the phase space). An additional integral of motion in involution
with $H_0(m,a)$, assuring the complete integrability of the system 
(\ref{LT rest}), is:
\begin{equation}\label{m3 rest}
m_3=\langle m,p\rangle \;.
\end{equation}

In the main text we give a Lagrangian derivation of equations of 
motion (\ref{LT body}) and (\ref{LT rest}) and an explanation of their 
Hamiltonian nature and integrability. Then we present a discrete Lagrangian 
function generating two maps approximating (\ref{LT body}) and (\ref{LT rest}), 
respectively. Most beautiful is the discretization of (\ref{LT rest}):
\begin{equation}\label{dLT rest}
\left\{\begin{array}{l}
m_{k+1}-m_k=\varepsilon\,p\times a_k\;,\\ \\ 
a_{k+1}-a_k=\displaystyle\frac{\varepsilon}{2}\,m_{k+1}\times 
(a_k+a_{k+1})\;.\end{array}
\right.
\end{equation}
It is easy to see that the second equation in (\ref{dLT rest}) can be uniquely
solved for $a_{k+1}$, so that (\ref{dLT rest}) defines a map $(m_k,a_k)\mapsto
(m_{k+1},a_{k+1})$, approximating, for small $\varepsilon$, the time 
$\varepsilon$ shift along the trajectories of (\ref{LT rest}). This 
distinguishes the 
situation from the one in \cite{MV} where Lagrangian equations led to 
correspondences rather than to maps. We shall demonstrate that the map
(\ref{dLT rest}) is Poisson with respect to the bracket (\ref{e3 br rest}),
so that the Casimirs (\ref{e3 Cas rest}) are integrals of motion. It is
also obvious that (\ref{m3 rest}) is an integral of motion. Most remarkably,
this map has another integral of motion -- an analog of the 
Hamiltonian:
\begin{equation}\label{H eps rest}
H_{\varepsilon}(m,a)=\frac{1}{2}\langle m,m\rangle +\langle a,p\rangle
+\frac{\varepsilon}{2}\langle a\times m,p\rangle\;.
\end{equation} 
The function (\ref{H eps rest}) is in involution with (\ref{m3 rest}), which 
renders the map (\ref{dLT rest}) completely integrable. A similar discretization
for the equations of motion in the body frame (\ref{LT body}) is slightly less
elegant.

\setcounter{equation}{0}
\section{Lagrangian mechanics on $TG$ \\ (continuous time case)}

Let $\bL(g,\dot{g})\,:\,TG\mapsto{\Bbb R}$ be a smooth function on the tangent
bundle of the Lie group $G$, called the {\it Lagrange function}. For an 
arbitrary function $g(t)\,:\,[t_0,t_1]\mapsto G$ one can consider the {\it 
action functional}
\begin{equation}\label{action}
S=\int_{t_0}^{t_1}\bL(g(t),\dot{g}(t))dt\;.
\end{equation} 
A standard argument shows that the functions $g(t)$ yielding extrema of this
functional (in the class of variations preserving $g(t_0)$ and $g(t_1)$), 
satisfy with necessity the {\it Euler--Lagrange equations}. In local coordinates
$\{g^i\}$ on $G$ they read:
\begin{equation}\label{EL gen}
\frac{d}{dt}\left(\frac{\partial\bL}{\partial\dot{g}^i}\right)=
\frac{\partial\bL}{\partial g^i}\;.
\end{equation}
The action functional $S$ is independent of the choice of local coordinates,
and thus the Euler--Lagrange equations are actually coordinate independent as
well. For a coordinate--free description in the language of differential
geometry, see \cite{A}, \cite{MR}.

Introducing the quantities \footnote{For the 
notations from the Lie groups theory used in this and subsequent sections see 
Appendix A.}
\begin{equation}\label{Pi}
\Pi=\nabla_{\dot{g}}\bL\in T_g^* G\;,
\end{equation}
one defines the {\it Legendre transformation}:
\begin{equation}\label{Legendre gen}
(g,\dot{g})\in TG\mapsto (g,\Pi)\in T^*G\;.
\end{equation}
If it is invertible, i.e. if $\dot{g}$ can be expressed through $(g,\Pi)$, 
then the the Legendre transformation of the Euler--Lagrange equations 
(\ref{EL gen}) yield  a {\it Hamiltonian system} on $T^*G$ with respect to the 
standard symplectic structure on $T^*G$ and with the Hamilton function
\begin{equation}\label{Ham gen}
H(g,\Pi)=\langle \Pi,\dot{g}\rangle-\bL(g,\dot{g})\;,
\end{equation}
(where, of course, $\dot{g}$ has to be expressed through $(g,\Pi)$). Finally,
we want to mention the Noether construction for deriving the existence of 
integrals of motion of the Euler--Lagrange equations from the symmetry groups
of the Lagrange function. We shall formulate the simplest form of 
Noether's theorem, where Lagrangian functions are invariant under the action of 
one--dimensional groups. Let $\zeta\in\gog$ be a fixed element, and consider 
a one--parameter subgroup 
\begin{equation}\label{G(zeta)}
G^{(\zeta)}=\{e^{c\zeta}\,:\,c\in{\Bbb R}\}\subset G\;.
\end{equation}
\begin{proposition} 
\begin{enumerate}
\item[{\rm a)}] Let the Lagrange function be invariant under the action of 
$G^{(\zeta)}$ on $TG$ induced by {\em left} translations on $G$:
\begin{equation}\label{Lagr left inv}
\bL(e^{c\zeta}g,L_{e^{c\zeta}*}\dot{g})=\bL(g,\dot{g})\;.
\end{equation}
Then the following function is an integral of motion of the Euler--Lagrange
equations:
\begin{equation}\label{Noether int left inv}
\langle R_g^*(\nabla_{\dot{g}}\bL),\zeta\rangle=\langle R_g^*\Pi,\zeta\rangle\;.
\end{equation}
\item[{\rm b)}] Let the Lagrange function be invariant under the action of 
$G^{(\zeta)}$ on $TG$ induced by {\em right} translations on $G$:
\begin{equation}\label{Lagr right inv}
\bL(ge^{c\zeta},R_{e^{c\zeta}*}\dot{g})=\bL(g,\dot{g})\;.
\end{equation}
Then the following function is an integral of motion of the Euler--Lagrange
equations:
\begin{equation}\label{Noether int right inv}
\langle L_g^*(\nabla_{\dot{g}}\bL),\zeta\rangle=\langle L_g^*\Pi,\zeta\rangle\;.
\end{equation}
\end{enumerate}
\end{proposition}
{\bf Proof.} Differentiate (\ref{Lagr left inv}) (or (\ref{Lagr right inv}))
with respect to $c$, set $c=0$, and use the Euler--Lagrange equations. \qed

For a detailed proof of the general version of the Noether theorem see \cite{A},
\cite{MR}.
\vspace{2mm}

In practice, it turns out to be more convenient to work not with the tangent
bundle $TG$, but with its trivializations $G\times\gog$, which is achieved by
translating the vector $\dot{g}\in T_g G$ into the group unit via left or right
translations.

\subsection{Left trivialization}
Consider the trivialization map
\begin{equation}\label{left triv}
(g,\Omega)\in G\times\gog\mapsto (g,\dot{g})\in TG\;,
\end{equation}
where
\begin{equation}\label{Omega}
\dot{g}=L_{g*}\Omega \quad\Leftrightarrow\quad \Omega=L_{g^{-1}*}\dot{g}\;.
\end{equation}
The trivialization (\ref{left triv}) of the tangent bundle $TG$ induces
the following trivialization of the cotangent bundle $T^*G$:
\begin{equation}\label{left triv *}
(g,M)\in G\times\gog^*\mapsto (g,\Pi)\in T^*G\;,
\end{equation}
where 
\begin{equation}\label{M}
\Pi=L_{g^{-1}}^* M \quad\Leftrightarrow\quad M=L_g^*\Pi\;.
\end{equation}

Denote the pull--back of the Lagrange function through
\begin{equation}\label{Lagr left}
\bL^{(l)}(g,\Omega)=\bL(g,\dot{g})\;,
\end{equation}
so that 
\[
\bL^{(l)}(g,\Omega)\,:\, G\times\gog\mapsto{\Bbb R}\;.
\]
We want to find differential equations satisfied by those functions
$(g(t),\Omega(t))\,:\,[t_0,t_1]\mapsto G\times\gog$ delivering extrema of
the action functional
\[
S^{(l)}=\int_{t_0}^{t_1}\bL^{(l)}(g(t),\Omega(t))dt
\]
and such that
\[
\Omega(t)=L_{g^{-1}(t)*}\dot{g}(t)\;.
\]
Admissible variations of $(g(t),\Omega(t))$ are those preserving the latter
equality and the values $g(t_0)$, $g(t_1)$.
\begin{proposition}
The differential equations for extremals of the functional $S^{(l)}$ read:
\begin{equation}\label{EL left Ham}
\left\{\begin{array}{l}
\dot{M}={\rm ad}^*\,\Omega\cdot M+d\,'_{\!g}\bL^{(l)}\;,\\ \\
\dot{g}=L_{g*}\Omega\;,\end{array}\right.
\end{equation}
where
\begin{equation}\label{M thru L}
M=\nabla_{\Omega}\bL^{(l)}\in\gog^*\;.
\end{equation}
If the Legendre transformation
\begin{equation}\label{Legendre left}
(g,\Omega)\in G\times\gog\mapsto (g,M)\in G\times\gog^*
\end{equation}
is invertible, it turns {\rm(\ref{EL left Ham})} into a Hamiltonian 
form on $G\times\gog^*$ with the Hamilton function
\begin{equation}\label{Ham left}
H(g,M)=\langle M,\Omega\rangle -\bL^{(l)}(g,\Omega)\;,
\end{equation}
(where, of course, $\Omega$ has to be expressed through $(g,M)$); the 
underlying invariant Poisson bracket on $G\times\gog^*$ is the pull--back 
of the standard symplectic bracket on $T^*G$, so that for two arbitrary 
functions $f_{1,2}(g,M)\,:\,G\times\gog^*\mapsto{\Bbb R}$ we have:
\begin{equation}\label{PB left}
\{f_1,f_2\}=-\langle d\,'_{\!g}f_1,\nabla_M f_2\rangle+
\langle d\,'_{\!g}f_2,\nabla_M f_1\rangle+
\langle M,[\nabla_M f_1,\nabla_M f_2]\,\rangle\;.
\end{equation}
\end{proposition}
{\bf Proof.} The equations of motion (\ref{EL left Ham}) may be derived by 
pulling back the equations (\ref{EL gen}) under the trivialization map 
(\ref{left triv}), but it is somewhat simpler to derive them independently. 
To this end, consider the admissible variations of $(g(t),\Omega(t))$ in the 
form
\[
g(t,\epsilon)=g(t)e^{\epsilon\eta(t)}\;,\quad{\rm where}\quad 
\eta(t)\,:\,[t_0,t_1]\mapsto\gog\;,\quad \eta(t_0)=\eta(t_1)=0\;,
\]
and
\begin{eqnarray*}
\Omega(t,\epsilon) & = & L_{g^{-1}(t,\epsilon)*}\dot{g}(t,\epsilon)=
{\rm Ad}\,e^{-\epsilon\eta(t)}\cdot \Omega(t)+\epsilon\dot{\eta}(t) 
+O(\epsilon^2)\\
 & = & \Omega(t)+\epsilon\Big(\dot{\eta}(t)+[\Omega(t),\eta(t)]\Big)
       +O(\epsilon^2)\;.
\end{eqnarray*}
So, equating the variation of action to zero, we get:
\[
0=\left.\frac{dS^{(l)}}{d\epsilon}\right|_{\epsilon=0}=
\int_{t_0}^{t_1}\Big(\langle\, d\,'_{\!g}\bL^{(l)},\eta\,\rangle+
\langle\nabla_{\Omega}\bL^{(l)},\dot{\eta}+{\rm ad}\,\Omega\cdot\eta\,\rangle
\Big)\,dt\;.
\]
Integrating the term with $\dot{\eta}$ by parts and taking into account
$\eta(t_0)=\eta(t_1)=0$, we come to:
\[
\int_{t_0}^{t_1}\langle\, d\,'_{\!g}\bL^{(l)}+{\rm ad}^*\,\Omega\cdot
\nabla_{\Omega}\bL^{(l)}-\frac{d}{dt}(\nabla_{\Omega}\bL^{(l)}),\eta\,\rangle
\,dt=0\;.
\]
Due to arbitariness of $\eta(t)$ the following equation holds:
\[
\frac{d}{dt}(\nabla_{\Omega}\bL^{(l)})=
{\rm ad}^*\,\Omega\cdot\nabla_{\Omega}\bL^{(l)}+d\,'_{\!g}\bL^{(l)}\;.
\] 
It remains to notice that $M$ defined by (\ref{M}), (\ref{Pi}), i.e. 
$M=L_g^*\nabla_{\dot{g}}\bL$, coincides with (\ref{M thru L}), as it follows
from (\ref{Omega}).  \qed
\vspace{2mm}

We now observe, what does Noether's theorem (more exactly, its version in 
Proposition 2.1) yield under left trivialization. 
\begin{proposition}
\begin{enumerate}
\item[{\rm a)}] Let the Lagrange function $\bL^{(l)}(g,\Omega)$ be invariant 
under the action of $G^{(\zeta)}$ on $G\times\gog$ induced by {\em left} 
translations on $G$:
\begin{equation}\label{left Lagr left inv}
\bL^{(l)}(e^{c\zeta}g,\Omega)=\bL^{(l)}(g,\Omega)\;.
\end{equation}
Then the following function is an integral of motion of the Euler--Lagrange
equations:
\begin{equation}\label{left int left inv}
\langle {\rm Ad}^*\,g^{-1}\cdot\nabla_{\Omega}\bL^{(l)},\zeta\rangle=
\langle M,{\rm Ad}\,g^{-1}\cdot\zeta\rangle\;.
\end{equation}
\item[{\rm b)}] Let the Lagrange function $\bL^{(l)}(g,\Omega)$ be invariant 
under the action of $G^{(\zeta)}$ on $G\times\gog$ induced by {\em right} 
translations on $G$:
\begin{equation}\label{left Lagr right inv}
\bL^{(l)}(ge^{c\zeta}, {\rm Ad}\,e^{-c\zeta}\cdot\Omega)=\bL^{(l)}(g,\Omega)\;.
\end{equation}
Then the following function is an integral of motion of the Euler--Lagrange
equations:
\begin{equation}\label{left int right inv}
\langle \nabla_{\Omega}\bL^{(l)},\zeta\rangle=\langle M,\zeta\rangle\;.
\end{equation}
\end{enumerate}
\end{proposition}

We finish this subsection by discussing the reduction procedure relevant for 
later applications. Let us assume that there holds a condition somewhat 
stronger than (\ref{left Lagr left inv}), namely, that the function $\bL^{(l)}$
is left invariant under the action of a subgroup somewhat larger than 
$G^{(\zeta)}$. Fix an element $\zeta\in\gog$, and consider the isotropy 
subgroup $G^{[\zeta]}$ of $\zeta$ with respect to the adjoint action 
of $G$, i.e.
\begin{equation}\label{G[zeta]}
G^{[\zeta]}=\{h\,:\;{\rm Ad}\,h\cdot\zeta=\zeta\}\subset G \;.
\end{equation}  
Obviously, $G^{(\zeta)}\subset G^{[\zeta]}$. Suppose that the Lagrange function 
$\bL^{(l)}(g,\Omega)$ is invariant under the action of $G^{[\zeta]}$ on 
$G\times\gog$ induced by {\it left} translations on $G$:
\begin{equation}\label{left action for red}
\bL^{(l)}(hg,\Omega)=\bL^{(l)}(g,\Omega)\;, \quad h\in G^{[\zeta]}\;.
\end{equation}
We want to reduce the Euler--Lagrange equations with respect to this action.
As a section $(G\times\gog)/G^{[\zeta]}$ we choose the set $\gog_{\zeta}\times
\gog$, where $\gog_{\zeta}$ is the orbit of $\zeta$ under the adjoint action
of $G$:
\begin{equation}\label{orbit}
\gog_{\zeta}=\{{\rm Ad}\,g\cdot\zeta\,,\;g\in G\}\subset
\gog\;.
\end{equation} 
We define the reduced Lagrange function $\cL^{(l)}\,:\,\gog_{\zeta}\times\gog
\mapsto{\Bbb R}$ as
\begin{equation}\label{left red Lagr}
\cL^{(l)}(P,\Omega)=\bL^{(l)}(g,\Omega)\;,\quad{\rm where}\quad
 P={\rm Ad}\,g^{-1}\cdot\zeta\;.
\end{equation}
This definition is correct, because from
\[
P={\rm Ad}\,g_1^{-1}\cdot\zeta={\rm Ad}\,g_2^{-1}\cdot\zeta
\]
there follows ${\rm Ad}\,g_2g_1^{-1}\cdot\zeta=\zeta$, so that $g_2g_1^{-1}\in
G^{[\zeta]}$, and $\bL^{(l)}(g_1,\Omega)=\bL^{(l)}(g_2,\Omega)$.
\begin{proposition}
Consider the reduction $(g,\Omega)\mapsto(P,\Omega)$. The reduced 
Euler--Lagrange equations {\rm (\ref{EL left Ham})} read:
\begin{equation}\label{EL left red Ham}
\left\{\begin{array}{l}
\dot{M}={\rm ad}^*\,\Omega\cdot M+{\rm ad}^*\,P\cdot\nabla_P\cL^{(l)}\;,\\ \\
\dot{P}=[P,\Omega]\;,\end{array}\right.
\end{equation}
where
\begin{equation}\label{M thru L red}
M=\nabla_{\Omega}\cL^{(l)}\in\gog^*\;.
\end{equation}
If the Legendre transformation
\begin{equation}\label{Legendre left red}
(P,\Omega)\in \gog_{\zeta}\times\gog\mapsto (P,M)\in \gog_{\zeta}\times\gog^*
\end{equation}
is invertible, it turns {\rm(\ref{EL left red Ham})} into a Hamiltonian
system on $\gog_{\zeta}\times\gog^*$, with the Hamilton function
\begin{equation}\label{Ham left red}
H(P,M)=\langle M,\Omega\rangle -\cL^{(l)}(P,\Omega)\;,
\end{equation}
where $\Omega$ has to be expressed through $(P,M)$; the underlying invariant
Poisson structure on $\gog_{\zeta}\times\gog^*$ is given by the following
formula: 
\begin{equation}\label{PB left red}
\{F_1,F_2\}=-\langle \nabla_P F_1,[P,\nabla_M F_2]\,\rangle+
\langle \nabla_P F_2,[P,\nabla_M F_1]\,\rangle+
\langle M,[\nabla_M F_1,\nabla_M F_2]\,\rangle
\end{equation}
for two arbitrary functions 
$F_{1,2}(P,M):\gog_{\zeta}\times\gog^*\mapsto{\Bbb R}$. (This formula indeed
defines a Poisson bracket on all of $\gog\times\gog^*$).

In addition to the integral of motion {\rm(\ref{Ham left red})}, the equations
of motion {\rm(\ref{EL left red Ham})} always have the following integral of 
motion:
\begin{equation}\label{integral of left red EL}
C=\langle\,M,P\,\rangle\;.
\end{equation}
This function is a Casimir of the bracket {\rm(\ref{PB left red})}.
\end{proposition}
{\bf Proof} is a consequence of the following formulas:
\[
d\,'_{\!g}\bL^{(l)}={\rm ad}^*\,P\cdot\nabla_P\cL^{(l)}\;,\qquad
\nabla_{\Omega}\bL^{(l)}=\nabla_{\Omega}\cL^{(l)}\;,
\]
which are easy to derive from the definitions, and similar formulas connecting
the Lie derivatives of $f_{1,2}$ with the gradients of $F_{1,2}$. \qed

\subsection{Right trivialization}
All constructions in this subsection are absolutely parallel to those of
the previous one, therefore we restrict ourselves to the formulation and
omit all proofs.

Consider the trivialization map
\begin{equation}\label{right triv}
(g,\omega)\in G\times\gog\mapsto (g,\dot{g})\in TG\;,
\end{equation}
where
\begin{equation}\label{omega}
\dot{g}=R_{g*}\omega \quad\Leftrightarrow\quad \omega=R_{g^{-1}*}\dot{g}\;.
\end{equation}
This trivialization of the tangent bundle $TG$ induces
the following trivialization of the cotangent bundle $T^*G$:
\begin{equation}\label{right triv *}
(g,m)\in G\times\gog^*\mapsto (g,\Pi)\in T^*G\;,
\end{equation}
where 
\begin{equation}\label{m}
\Pi=R_{g^{-1}}^* m \quad\Leftrightarrow\quad m=R_g^*\Pi\;.
\end{equation}
The pull--back of the Lagrange function is denoted through
\begin{equation}\label{Lagr right}
\bL^{(r)}(g,\omega)=\bL(g,\dot{g})\;.
\end{equation}
\begin{proposition}
The differential equations for the extremals of the functional 
\[
S^{(r)}=\int_{t_0}^{t_1}\bL^{(r)}(g(t),\omega(t))dt
\] 
read:
\begin{equation}\label{EL right Ham}
\left\{\begin{array}{l}
\dot{m}=-{\rm ad}^*\,\omega\cdot m + d_g\bL^{(r)}\;,\\ \\
\dot{g}=R_{g*}\omega\;,\end{array}\right.
\end{equation}
where
\begin{equation}\label{m thru L}
m=\nabla_{\omega}\bL^{(r)}\in\gog^*\;.
\end{equation}
If the Legendre transformation
\begin{equation}\label{Legendre right}
(g,\omega)\in G\times\gog\mapsto (g,m)\in G\times\gog^*
\end{equation}
is invertible, it turns {\rm(\ref{EL right Ham})} into a Hamiltonian
form on $G\times\gog^*$ with the Hamilton function
\begin{equation}\label{Ham right}
H(g,m)=\langle m,\omega\rangle -\bL^{(r)}(g,\omega)\;,
\end{equation}
where $\omega$ has to be expressed through $(g,m)$; the underlying invariant
Poisson bracket on $G\times\gog^*$ is the pull--back of the standard symplectic 
bracket on $T^*G$, so that for two arbitrary functions 
$f_{1,2}(g,m)\,:\,G\times\gog^*\mapsto{\Bbb R}$ we have:
\begin{equation}\label{PB right}
\{f_1,f_2\}=-\langle d_g f_1,\nabla_m f_2\rangle+
\langle d_g f_2,\nabla_m f_1\rangle-
\langle m,[\nabla_m f_1,\nabla_m f_2]\,\rangle\;.
\end{equation}
\end{proposition}

A version of Noether's theorem takes the following form:
\begin{proposition}
\begin{enumerate}
\item[{\rm a)}] Let the Lagrange function $\bL^{(r)}(g,\omega)$ be invariant 
under the action of $G^{(\zeta)}$ on $G\times\gog$ induced by {\em left} 
translations on $G$:
\begin{equation}\label{right Lagr left inv}
\bL^{(r)}(e^{c\zeta}g, {\rm Ad}\,e^{c\zeta}\cdot\omega)=\bL^{(r)}(g,\omega)\;.
\end{equation}
Then the following function is an integral of motion of the Euler--Lagrange
equations:
\begin{equation}\label{right int left inv}
\langle \nabla_{\omega}\bL^{(r)},\zeta\rangle=\langle m,\zeta\rangle\;.
\end{equation}
\item[{\rm b)}] Let the Lagrange function $\bL^{(r)}(g,\omega)$ be invariant 
under the action of $G^{(\zeta)}$ on $G\times\gog$ induced by {\em right} 
translations on $G$:
\begin{equation}\label{right Lagr right inv}
\bL^{(r)}(ge^{c\zeta}, \omega)=\bL^{(r)}(g,\omega)\;.
\end{equation}
Then the following function is an integral of motion of the Euler--Lagrange
equations:
\begin{equation}\label{right int right inv}
\langle {\rm Ad}^*\,g\cdot\nabla_{\omega}\bL^{(r)},\zeta\rangle=
\langle m,{\rm Ad}\,g\cdot\zeta\rangle\;.
\end{equation}
\end{enumerate}
\end{proposition}

Turning to the reduction procedure, suppose that the Lagrange function 
$\bL^{(r)}(g,\omega)$ is invariant under the action of $G^{[\zeta]}$ on 
$G\times\gog$ induced by {\it right} translations on $G$:
\begin{equation}\label{right action for red}
\bL^{(r)}(gh,\omega)=\bL^{(r)}(g,\omega)\;, \quad h\in G^{[\zeta]}
\end{equation}
We define the reduced Lagrange function $\cL^{(r)}\,:\,\gog_{\zeta}\times\gog
\mapsto{\Bbb R}$ as
\begin{equation}\label{right red Lagr}
\cL^{(r)}(a,\omega)=\bL^{(r)}(g,\omega)\;,\quad{\rm where}\quad
 a={\rm Ad}\,g\cdot\zeta\;.
\end{equation}
\begin{proposition}
Consider the reduction $(g,\omega)\mapsto(a,\omega)$. The reduced 
Euler--Lagrange equations {\rm (\ref{EL right Ham})} read:
\begin{equation}\label{EL right red Ham}
\left\{\begin{array}{l}
\dot{m}=-{\rm ad}^*\,\omega\cdot m-{\rm ad}^*\,a\cdot\nabla_a\cL^{(r)}\;,\\ \\
\dot{a}=[\omega,a]\;,\end{array}\right.
\end{equation}
where 
\begin{equation}\label{m thru L red}
m=\nabla_{\omega}\cL^{(r)}\in\gog^*\;.
\end{equation}
If the Legendre transformation
\begin{equation}\label{Legendre right red}
(a,\omega)\in \gog_{\zeta}\times\gog\mapsto (a,m)\in \gog_{\zeta}\times\gog^*
\end{equation}
is invertible, it turns {\rm(\ref{EL right red Ham})} into a Hamiltonian
system on $\gog_{\zeta}\times\gog^*$, with the Hamilton function
\begin{equation}\label{Ham right red}
H(a,m)=\langle m,\omega\rangle -\cL^{(r)}(a,\omega)\;,
\end{equation}
where $\omega$ has to be expressed through $(a,m)$; the underlying invariant
Poisson structure on $\gog_{\zeta}\times\gog^*$ is given by the following
formula:
\begin{equation}\label{PB right red}
\{F_1,F_2\}=\langle \nabla_a F_1,[a,\nabla_m F_2]\,\rangle-
\langle \nabla_a F_2,[a,\nabla_m F_1]\,\rangle-
\langle m,[\nabla_m F_1,\nabla_m F_2]\,\rangle
\end{equation}
for two arbitrary functions 
$F_{1,2}(a,m):\gog_{\zeta}\times\gog^*\mapsto{\Bbb R}$. (This formula indeed
defines a Poisson bracket on all of $\gog\times\gog^*$).

In addition to the integral of motion {\rm(\ref{Ham right red})}, the equations
of motion {\rm(\ref{EL right red Ham})} always have the following integral of 
motion:
\begin{equation}\label{integral of right red EL}
c=\langle\,m,a\,\rangle\;.
\end{equation}
This function is a Casimir of the bracket {\rm(\ref{PB right red})}.
\end{proposition}
Notice that the brackets (\ref{PB left red}) and (\ref{PB right red})
essentially coincide (differ only by a sign).
\vspace{2mm}

{\bf Remark.} For future reference notice that the elements $\Omega,\omega\in
\gog$ and $M,m\in\gog^*$ are related via the formulas
\begin{equation}\label{Omega omega}
\Omega={\rm Ad}\,g^{-1}\cdot\omega\;,
\end{equation}
\begin{equation}\label{M m}
M={\rm Ad}^*\,g\cdot m\;.
\end{equation}

\setcounter{equation}{0}
\section{Lagrangian mechanics on $G\times G$ \\ (discrete time case)}

We now turn to the discrete time analog of the previous constructions,
introduced in \cite{V}, \cite{MV}. Our presentation is an adaptation of
the Moser--Veselov construction for the case when the basic manifold is a 
Lie group. We shall see that almost all constructions of the previous section
have their discrete time analogs. The only exception is the existence of the
``energy'' integral (\ref{Ham gen}).

Let $\bbL(g,\wg)\,:\,G\times G$ be a smooth function, called the (discrete
time) {\it Lagrange function}. For an arbitrary sequence $\{g_k\in G,\;k=k_0,
k_0+1,\ldots,k_1\}$ one can consider the {\it action functional}
\begin{equation}\label{dS}
S=\sum_{k=k_0}^{k_1-1}\bbL(g_k,g_{k+1})\;.
\end{equation}
Obviously, the sequences $\{g_k\}$ delivering extrema of this functional
(in the class of variations preserving $g_{k_0}$ and $g_{k_1}$), satisfy
with necessity the {\it discrete Euler--Lagrange equations}:
\begin{equation}\label{dEL}
\nabla_1\bbL(g_k,g_{k+1})+\nabla_2\bbL(g_{k-1},g_k)=0\;.
\end{equation}
Here $\nabla_1\bbL(g,\wg)$ ($\nabla_2\bbL(g,\wg)$) denotes the gradient of 
$\bbL(g,\wg)$ with respect to the first argument $g$ (resp. the second argument
$\wg$). Notice that in our case, when $G$ is a Lie group and not just a general
smooth manifold, the equation (\ref{dEL}) is written in a coordinate free 
form, using the intrinsic notions of the Lie theory. This is opposed to the
continuous time case and somehow underlines the fundamental character of the
discrete Euler--Lagrange equations. 

The equation (\ref{dEL}) is an implicit equation for $g_{k+1}$. In 
general, it has more than one solution, and therefore defines a correspondence
(multi--valued map)  $(g_{k-1},g_k)\mapsto(g_k,g_{k+1})$. To discuss symplectic
properties of this correspondence, one defines:
\begin{equation}\label{dPi}
\Pi_k=\nabla_2\bbL(g_{k-1},g_k)\in T^*_{g_k}G\;.
\end{equation} 
Then (\ref{dEL}) may be rewritten as the following system:
\begin{equation}\label{dEL syst}
\left\{\begin{array}{l}
\Pi_k=-\nabla_1\bbL(g_k,g_{k+1}) \\ \\ \Pi_{k+1}=\nabla_2\bbL(g_k,g_{k+1}) 
\end{array}\right.
\end{equation}
This system defines a (multivalued) map $(g_k,\Pi_k)\mapsto(g_{k+1},\Pi_{k+1})$
of $T^*G$ into itself. More precisely, the first equation in (\ref{dEL syst})
is an implicit equation for $g_{k+1}$, while the second one allows for the
explicit and unique calculation of $\Pi_{k+1}$, knowing $g_k$ and $g_{k+1}$. As 
demonstrated in \cite{V}, \cite{MV}, this map $T^*G\mapsto T^*G$ is symplectic
with respect to the standard symplectic structure on $T^*G$.

For discrete Euler--Lagrange equations there holds an analog of Noether's
theorem. Again, we give only the simplest version thereof.
\begin{proposition} 
\begin{enumerate}
\item[{\rm a)}] Let the Lagrange function be invariant under the action of 
$G^{(\zeta)}$ on $G\times G$ induced by {\em left} translations on $G$:
\begin{equation}\label{dLagr left inv}
\bbL(e^{c\zeta}g,e^{c\zeta}\wg)=\bbL(g,\wg)\;.
\end{equation}
Then the following function is an integral of motion of the discrete 
Euler--Lagrange equations:
\begin{equation}\label{dNoether int left inv}
\langle d_2\bbL(g_{k-1},g_k),\zeta\rangle=\langle R_{g_k}^*\Pi_k,\zeta\rangle\;.
\end{equation}
\item[{\rm b)}] Let the Lagrange function be invariant under the action of 
$G^{(\zeta)}$ on $G\times G$ induced by {\em right} translations on $G$:
\begin{equation}\label{dLagr right inv}
\bbL(ge^{c\zeta},\wg e^{c\zeta})=\bbL(g,\wg)\;.
\end{equation}
Then the following function is an integral of motion of the Euler--Lagrange
equations:
\begin{equation}\label{dNoether int right inv}
\langle -d\,'_{\!1}\bbL(g_k,g_{k+1}),\zeta\rangle=\langle L_{g_k}^*\Pi_k,
\zeta\rangle\;.
\end{equation}
\end{enumerate}
\end{proposition}
{\bf Proof.} Since both statements are proved similarly, we restrict ourselves 
to proof of the first one. To this end differentiate (\ref{dLagr left inv})
with respect to $c$ and set $c=0$. Writing $(g_k,g_{k+1})$ for $(g,\wg)$, we
get:
\[
\langle d_1\bbL(g_k,g_{k+1}),\zeta\rangle + 
\langle d_2\bbL(g_k,g_{k+1}),\zeta\rangle =0\;.
\]
But the discrete Euler--Lagrange equations imply that
\[
d_1\bbL(g_k,g_{k+1})=-d_2\bbL(g_{k-1},g_k)\;.
\]
Hence
\[
\langle d_2\bbL(g_k,g_{k+1}),\zeta\rangle=
\langle d_2\bbL(g_{k-1},g_k),\zeta\rangle\;,
\]
and the statement is proved. \qed

Notice that the expressions of the Noether integrals in terms of $(g,\Pi)$ are
{\it exactly} the same as in the continuous time case.

\subsection{Left trivialization}
Actually, the tangent bundle $TG$ does not appear in the discrete time context
at all. We shall see that the analogs of the ``angular velocities'' $\Omega$,
$\omega$ live not in $T_gG$ but in $G$ itself. On the contrary, the cotangent 
bundle $T^*G$ still plays an important role in the discrete time theory, and it
is still convenient to trivialize it. This subsection is devoted to the 
constructions related to the left trivialization.

Consider the map
\begin{equation}\label{d left triv}
(g_k,W_k)\in G\times G \mapsto (g_k,g_{k+1})\in G\times G\;,
\end{equation}
where
\begin{equation}\label{W}
g_{k+1}=g_kW_k \quad\Leftrightarrow\quad W_k=g_k^{-1}g_{k+1}\;.
\end{equation}
The group element $W_k$ is an analog of the left angular velocity $\Omega$
from (\ref{Omega}), more precisely, it approximates $e^{\epsilon\Omega}$.
Consider also the left trivialization of the cotangent bundle $T^*G$:
\begin{equation}\label{d left triv *}
(g_k,M_k)\in G\times\gog^*\mapsto (g_k,\Pi_k)\in T^*G\;,
\end{equation}
where 
\begin{equation}\label{d M}
\Pi_k=L_{g_k^{-1}}^* M_k \quad\Leftrightarrow\quad M_k=L_{g_k}^*\Pi_k\;.
\end{equation}

Denote the pull--back of the Lagrange function under (\ref{d left triv}) through
\begin{equation}\label{dLagr left}
\bbL^{(l)}(g_k,W_k)=\bbL(g_k,g_{k+1})\;.
\end{equation}
We want to find difference equations satisfied by the sequences
$\{(g_k,W_k)\,,k=k_0,\ldots,k_1-1\}$ delivering extrema of the action 
functional
\[
S^{(l)}=\sum_{k_0}^{k_1-1}\bbL^{(l)}(g_k,W_k)\;,
\]
and satisfying $W_k=g_k^{-1}g_{k+1}$.
Admissible variations of $\{(g_k, W_k)\}$ are those preserving the values 
of $g_{k_0}$ and $g_{k_1}=g_{k_1-1}W_{k_1-1}$.
\begin{proposition}
The difference equations for extremals of the functional $S^{(l)}$ read:
\begin{equation}\label{dEL left Ham}
\left\{\begin{array}{l}
{\rm Ad}^*\,W_k^{-1}\cdot M_{k+1}=M_k+d\,'_{\!g}\bbL^{(l)}(g_k,W_k)\;,\\ \\
g_{k+1}=g_kW_k\;,\end{array}\right.
\end{equation}
where
\begin{equation}\label{d M thru L}
M_k=d\,'_{\!W}\bbL^{(l)}(g_{k-1},W_{k-1})\in\gog^*\;.
\end{equation}
If the ``Legendre transformation''
\begin{equation}\label{dLegendre left}
(g_{k-1},W_{k-1})\in G\times G\mapsto (g_k,M_k)\in G\times\gog^*\;,
\end{equation}
where $g_k=g_{k-1}W_{k-1}$, is invertible, then {\rm(\ref{dEL left Ham})} 
defines a map $(g_k,M_k)\mapsto(g_{k+1},M_{k+1})$ which is symplectic 
with respect to the Poisson bracket {\rm(\ref{PB left})} on $G\times\gog^*$.
\end{proposition}
{\bf Proof.} The simplest way to derive (\ref{dEL left Ham}) is to pull
back the equations (\ref{dEL}) under the map (\ref{d left triv}).
To do this, first rewrite (\ref{dEL}) as
\begin{equation}\label{dEL left aux1}
d\,'_{\!1}\bbL(g_k,g_{k+1})+d\,'_{\!2}\bbL(g_{k-1},g_k)=0\;.
\end{equation}
We have to express these Lie derivatives in terms of $(g,W)$. 
The answer is this:
\begin{equation}\label{dEL left aux2}
d\,'_{\!2}\bbL(g_{k-1},g_k)=d\,'_{\!W}\bbL^{(l)}(g_{k-1},W_{k-1})\;,
\end{equation}
\begin{equation}\label{dEL left aux3}
d\,'_{\!1}\bbL(g_k,g_{k+1})=d\,'_{\!g}\bbL^{(l)}(g_k,W_k)-
d_W\bbL^{(l)}(g_k,W_k)\;.
\end{equation}
Indeed, let us prove, for example, the (less obvious) (\ref{dEL left aux3}).
We have:
\begin{eqnarray*}
\langle d\,'_{\!1}\bbL(g_k,g_{k+1}),\eta\rangle & = & 
\left.\frac{d}{d\epsilon}\,
\bbL(g_ke^{\epsilon\eta},g_{k+1})\right|_{\epsilon=0}=
\left.\frac{d}{d\epsilon}\,
\bbL^{(l)}(g_ke^{\epsilon\eta},e^{-\epsilon\eta}W_k)\right|_{\epsilon=0}
\\ \\ & = & \langle d\,'_{\!g}\bbL^{(l)}(g_k,W_k),\eta\rangle-
\langle d_W\bbL^{(l)}(g_k,W_k),\eta\rangle\;.
\end{eqnarray*}
It remains to substitute (\ref{dEL left aux2}), (\ref{dEL left aux3}) into 
(\ref{dEL left aux1}). Taking into account that
\[
d_W\bbL^{(l)}(g_k,W_k)={\rm Ad}^*\,W_k^{-1}\cdot d\,'_{\!g}\bbL^{(l)}(g_k,W_k)
\]
we find (\ref{dEL left Ham}). Finally, notice that the notation 
(\ref{d M thru L}) is consistent with the definitions (\ref{dPi}), (\ref{d M}). 
Indeed, from these definitions it follows: $M_k=d\,'_{\!2}\bbL(g_{k-1},g_k)$, 
and the reference to (\ref{dEL left aux2}) finishes the proof. \qed
\vspace{2mm}

We now observe, what does the discrete time version of the Noether theorem 
from Proposition 3.1 yield under left trivialization. 
\begin{proposition}
\begin{enumerate}
\item[{\rm a)}] Let the Lagrange function $\bbL^{(l)}(g,W)$ be invariant 
under the action of $G^{(\zeta)}$ on $G\times\gog$ induced by {\em left} 
translations on $G$:
\begin{equation}\label{left dLagr left inv}
\bbL^{(l)}(e^{c\zeta}g,W)=\bbL^{(l)}(g,W)\;.
\end{equation}
Then the following function is an integral of motion of the Euler--Lagrange
equations:
\begin{equation}\label{d left int left inv}
\langle {\rm Ad}^*\,g_k^{-1}\cdot d\,'_{\!W}\bbL^{(l)}(g_{k-1},W_{k-1}),
\zeta\rangle = \langle M_k,{\rm Ad}\,g_k^{-1}\cdot\zeta\rangle\;.
\end{equation}
\item[{\rm b)}] Let the Lagrange function $\bbL^{(l)}(g,W)$ be invariant 
under the action of $G^{(\zeta)}$ on $G\times G$ induced by {\em right} 
translations on $G$:
\begin{equation}\label{left dLagr right inv}
\bbL^{(l)}(ge^{c\zeta}, e^{-c\zeta}We^{c\zeta})=\bbL^{(l)}(g,W)\;.
\end{equation}
Then the following function is an integral of motion of the Euler--Lagrange
equations:
\begin{equation}\label{d left int right inv}
\langle d\,'_{\!W}\bbL^{(l)}(g_{k-1},W_{k-1}),\zeta\rangle=
\langle M_k,\zeta\rangle\;.
\end{equation}
\end{enumerate}
\end{proposition}

We discuss now the reduction procedure. Assume that the function $\bbL^{(l)}$
is invariant under the action of $G^{[\zeta]}$ on 
$G\times G$ induced by {\it left} translations on $G$:
\begin{equation}\label{d left action for red}
\bbL^{(l)}(hg,W)=\bbL^{(l)}(g,W)\;, \quad h\in G^{[\zeta]}\;.
\end{equation}
Define the reduced Lagrange function $\Lambda^{(l)}\,:\,\gog_{\zeta}\times G
\mapsto{\Bbb R}$ as
\begin{equation}\label{left red dLagr}
\Lambda^{(l)}(P,W)=\bbL^{(l)}(g,W)\;,\quad{\rm where}\quad
 P={\rm Ad}\,g^{-1}\cdot\zeta\;.
\end{equation}
\begin{proposition}
Consider the reduction $(g,W)\mapsto(P,W)$. The reduced Euler--Lagrange 
equations {\rm (\ref{dEL left Ham})} read:
\begin{equation}\label{dEL left Ham red}
\left\{\begin{array}{l}
{\rm Ad}^*\,W_k^{-1}\cdot M_{k+1}=M_k+{\rm ad}^*\,P_k\cdot\nabla_P
\Lambda^{(l)}(P_k,W_k)\;,\\ \\
P_{k+1}={\rm Ad}\,W_k^{-1}\cdot P_k\;,\end{array}\right.
\end{equation}
where
\begin{equation}\label{d M thru L red}
M_k=d\,'_{\!W}\Lambda^{(l)}(P_{k-1},W_{k-1})\in\gog^*\;.
\end{equation}
If the ``Legendre transformation''
\begin{equation}\label{dLegendre left red}
(P_{k-1},W_{k-1})\in \gog_{\zeta}\times G\mapsto 
(P_k,M_k)\in \gog_{\zeta}\times\gog^*\;,
\end{equation}
where $P_k={\rm Ad}\,W_{k-1}^{-1}\cdot P_{k-1}$, is invertible, then
{\rm(\ref{dEL left Ham red})} define a map $(P_k,M_k)\mapsto(P_{k+1},M_{k+1})$ 
of  $\gog_{\zeta}\times\gog^*$ which is Poisson with respect to the Poisson 
bracket {\rm(\ref{PB left red})}.

The equations of motion {\rm(\ref{dEL left Ham red})} always have the following 
integral of motion:
\begin{equation}\label{integral of left red dEL}
C=\langle\,M_k,P_k\,\rangle\;,
\end{equation}
which is a Casimir function of the bracket {\rm(\ref{PB left red})}.
\end{proposition}

\subsection{Right trivialization}

Consider the map
\begin{equation}\label{d right triv}
(g_k,w_k)\in G\times G \mapsto (g_k,g_{k+1})\in G\times G\;,
\end{equation}
where
\begin{equation}\label{w}
g_{k+1}=w_kg_k \quad\Leftrightarrow\quad w_k=g_{k+1}g_k^{-1}\;.
\end{equation}
Consider also the right trivialization of the cotangent bundle $T^*G$:
\begin{equation}\label{d right triv *}
(g_k,m_k)\in G\times\gog^*\mapsto (g_k,\Pi_k)\in T^*G\;,
\end{equation}
where 
\begin{equation}\label{d m}
\Pi_k=R_{g_k^{-1}}^* m_k \quad\Leftrightarrow\quad m_k=R_{g_k}^*\Pi_k\;.
\end{equation}

Denote the pull--back of the Lagrange function under (\ref{d right triv}) 
through
\begin{equation}\label{dLagr right}
\bbL^{(r)}(g_k,w_k)=\bbL(g_k,g_{k+1})\;.
\end{equation}
\begin{proposition}
The difference equations for extremals of the functional 
\[
S^{(r)}=\sum_{k_0}^{k_1-1}\bbL^{(r)}(g_k,w_k)\;,
\] 
read:
\begin{equation}\label{dEL right Ham}
\left\{\begin{array}{l}
{\rm Ad}^*\,w_k\cdot m_{k+1}=m_k+d_g\bbL^{(r)}(g_k,w_k)\;,\\ \\
g_{k+1}=w_kg_k\;,\end{array}\right.
\end{equation}
where
\begin{equation}\label{d m thru L}
m_k=d_w\bbL^{(r)}(g_{k-1},w_{k-1})\in\gog^*\;.
\end{equation}
If the ``Legendre transformation'' 
\begin{equation}\label{dLegendre right}
(g_{k-1},w_{k-1})\in G\times G\mapsto (g_k,m_k)\in G\times\gog^*\;,
\end{equation}
where $g_k=w_{k-1}g_{k-1}$, is invertible, then {\rm(\ref{dEL right Ham})} 
define a map $(g_k,m_k)\mapsto(g_{k+1},m_{k+1})$ which is symplectic with 
respect to the Poisson bracket {\rm(\ref{PB right})} on $G\times\gog^*$.
\end{proposition}
{\bf Proof.} This time the discrete Euler--Lagrange equations (\ref{dEL})
are rewritten  as
\begin{equation}\label{dEL right aux1}
d_1\bbL(g_k,g_{k+1})+d_2\bbL(g_{k-1},g_k)=0\;,
\end{equation}
and the expressions for these Lie derivatives in terms of $(g,w)$ read: 
\begin{equation}\label{dEL right aux2}
d_2\bbL(g_{k-1},g_k)=d_w\bbL^{(r)}(g_{k-1},w_{k-1})\;,
\end{equation}
\begin{equation}\label{dEL right aux3}
d_1\bbL(g_k,g_{k+1})=d_g\bbL^{(r)}(g_k,w_k)-
d\,'_{\!w}\bbL^{(r)}(g_k,w_k)=d_g\bbL^{(r)}(g_k,w_k)-
{\rm Ad}^*\,w_k\cdot d_w\bbL^{(r)}(g_k,w_k)\;.
\end{equation}
The expression (\ref{d m thru L}) is consistent with the definitions
(\ref{dPi}), (\ref{d m}), which imply that $m_k=d_2\bbL(g_{k-1},g_k)$,
and a reference to (\ref{dEL right aux2}) finishes the proof. \qed
\vspace{2mm}

\begin{proposition}
\begin{enumerate}
\item[{\rm a)}] Let the Lagrange function $\bbL^{(r)}(g,w)$ be invariant 
under the action of $G^{(\zeta)}$ on $G\times G$ induced by {\em left} 
translations on $G$:
\begin{equation}\label{right dLagr left inv}
\bbL^{(r)}(e^{c\zeta}g, e^{c\zeta}we^{-c\zeta})=\bbL^{(r)}(g,w)\;.
\end{equation}
Then the following function is an integral of motion of the Euler--Lagrange
equations:
\begin{equation}\label{d right int left inv}
\langle d_w\bbL^{(r)}(g_{k-1},w_{k-1}),\zeta\rangle=
\langle m_k,\zeta\rangle\;.
\end{equation}
\item[{\rm b)}] Let the Lagrange function $\bbL^{(r)}(g,w)$ be invariant 
under the action of $G^{(\zeta)}$ on $G\times\gog$ induced by {\em right} 
translations on $G$:
\begin{equation}\label{right dLagr right inv}
\bbL^{(r)}(ge^{c\zeta},w)=\bbL^{(r)}(g,w)\;.
\end{equation}
Then the following function is an integral of motion of the Euler--Lagrange
equations:
\begin{equation}\label{d right int right inv}
\langle {\rm Ad}^*\,g_k\cdot d_w\bbL^{(r)}(g_{k-1},w_{k-1}),
\zeta\rangle = \langle m_k,{\rm Ad}\,g_k\cdot\zeta\rangle\;.
\end{equation}
\end{enumerate}
\end{proposition}

Finally, we turn to the reduction procedure. Assume that the function 
$\bbL^{(r)}$ is invariant under the action of $G^{[\zeta]}$ on 
$G\times G$ induced by {\it right} translations on $G$:
\begin{equation}\label{d right action for red}
\bbL^{(r)}(gh,w)=\bbL^{(r)}(g,w)\;, \quad h\in G^{[\zeta]}\;.
\end{equation}
Define the reduced Lagrange function $\Lambda^{(r)}\,:\,\gog_{\zeta}\times G
\mapsto{\Bbb R}$ as
\begin{equation}\label{right red dLagr}
\Lambda^{(r)}(a,w)=\bbL^{(r)}(g,w)\;,\quad{\rm where}\quad
 a={\rm Ad}\,g\cdot\zeta\;.
\end{equation}
\begin{proposition}
Consider the reduction $(g,w)\mapsto(a,w)$. The reduced Euler--Lagrange 
equations {\rm (\ref{dEL right Ham})} read:
\begin{equation}\label{dEL right Ham red}
\left\{\begin{array}{l}
{\rm Ad}^*\,w_k\cdot m_{k+1}=m_k-{\rm ad}^*\,a_k\cdot\nabla_a
\Lambda^{(r)}(a_k,w_k)\;,\\ \\
a_{k+1}={\rm Ad}\,w_k\cdot a_k\;,\end{array}\right.
\end{equation}
where
\begin{equation}\label{d m thru L red}
m_k=d_w\Lambda^{(r)}(a_{k-1},w_{k-1})\in\gog^*\;.
\end{equation}
If the ``Legendre transformation'' 
\begin{equation}\label{dLegendre right red}
(a_{k-1},w_{k-1})\in \gog_{\zeta}\times G\mapsto 
(a_k,m_k)\in \gog_{\zeta}\times\gog^*\;,
\end{equation}
where $a_k={\rm Ad}\,w_{k-1}\cdot a_{k-1}$, is invertible, then {\rm(\ref
{dEL right Ham red})} define a map $(a_k,m_k)\mapsto(a_{k+1},m_{k+1})$ of  
$\gog_{\zeta}\times\gog^*$ which is Poisson with respect to the bracket 
{\rm(\ref{PB right red})}.

The equations of motion {\rm(\ref{dEL right Ham red})} always have the 
following integral of motion:
\begin{equation}\label{integral of right red dEL}
c=\langle\,m_k,a_k\,\rangle\;,
\end{equation}
which is a Casimir of the bracket {\rm(\ref{PB right red})}.
\end{proposition}

A table summarizing the unreduced and reduced Lagrangian equations of motion, 
both in the continuous and discrete time formulations, is put in Appendix B.

\setcounter{equation}{0}
\section{Lagrangian formulation of the Lagrange top}

From now on we always work with the group $G=SU(2)$, so that $\gog=su(2)$,
see Appendix C for necessary background. In particular, we identify vectors
from ${\Bbb R}^3$ with matrices from $\gog$, and do not distinguish between
the vector product in ${\Bbb R}^3$ and the commutator in $\gog$. We write the 
adjoint group action as a matrix conjugation, and the operators $L_g^*$, 
$R_g^*$ as left and right matrix multiplication by $g^{-1}$, in accordance with
(\ref{ops}) and (\ref{ops*}).

The following table summarizes the integrals of motion and the reductions 
following from the symmetries of Lagrange functions, in the terminology
of the rigid body motion.  
\vspace{0.7cm}

\begin{tabular}{|c||c|c|}\hline
 & Left symmetry  & Right symmetry  \\
 & $g\mapsto e^{cp}g$ & $g\mapsto ge^{cA}$ \\
 & (rotation about $p\,,$ & (rotation about $A\,,$ \\
 & the gravity field axis) & the body symmetry axis)\\
\hline\hline
Left trivialization  &  & \\
$(g,\Pi)\mapsto (g,M=g^{-1}\Pi)$ & $\langle M,P\rangle\,,\;P=g^{-1}pg$ & 
$\langle M,A\rangle$ \\
(body frame) & & \\
\hline
Right trivialization & & \\
$(g,\Pi)\mapsto (g,m=\Pi g^{-1})$ &$\langle m,p\rangle$ & 
$\langle m,a\rangle\,,\;a=gAg^{-1}$ \\
(rest frame)  & & \\
\hline
\end{tabular} 
\vspace{3mm}

\subsection{Body frame formulation}

For an arbitrary Lagrangian system on $TG$, whose Lagrange function
may be written as
\[
\bL(g,\dot{g})=\cL^{(l)}(P,\Omega)\;,
\]
where $\Omega=g^{-1}\dot{g}$, $P=g^{-1}pg$, the Euler--Lagrange equations
of motion take the form
\begin{equation}\label{Left inv Lagr}
\left\{\begin{array}{l} \dot{M}=[M,\Omega]+[\nabla_P \cL^{(l)},P]\;\\ \\
\dot{P}=[P,\Omega]\;,\end{array}\right.
\end{equation}
where $M=\nabla_{\Omega}\cL^{(l)}$. Such systems are characterized by the 
condition of invariance of $\bL(g,\dot{g})$ under the action of $G^{(p)}$ 
on $TG$ induced by left translations on $G$, i.e.
\[
\bL(e^{cp}g,e^{cp}\dot{g})=\bL(g,\dot{g})\;.
\]
The geometrical meaning of this action is the rotation around $p$ -- the 
symmetry axis of the gravitation field. 
Consider the Lagrange function of the general top:
\begin{equation}\label{Top Lagr}
\cL^{(l)}(P,\Omega)=\frac{1}{2}\,\langle J\Omega,\Omega\rangle
-\langle P,A\rangle\,,
\end{equation} 
where $J:\gog\mapsto\gog$ is a linear operator, and $A\in\gog$ is a constant
vector. We calculate:
\[
M=\nabla_{\Omega}\cL^{(l)}=J\Omega\;,\qquad \nabla_P\cL^{(l)}=-A\;,
\]
so that (\ref{Left inv Lagr}) takes the form
\begin{equation}\label{Top body}
\left\{\begin{array}{l} \dot{M}=[M,\Omega]+[P,A]\;,\\ \\
\dot{P}=[P,\Omega]\;,\end{array}\right.
\end{equation}
where
\begin{equation}\label{Top M}
M=J\Omega\;,
\end{equation}
which is identical with (\ref{LT body}). According to Proposition 2.4, this
system is Hamiltonian with respect to the bracket (\ref{PB left red}), which 
in our case has the coordinate representation (\ref{e3 br body}).

The Lagrange top is distinguished by the relations (\ref{A body}).
They may be represented in the following, slightly more invariant fashion:
\begin{equation}\label{LT inertia}
M=J\Omega=\Omega-(1-\alpha)\langle\Omega,A\rangle A\;,
\end{equation}
i.e. $J$ acts as multiplication by $\alpha$ in the direction of the vector  $A$, 
and as the identity operator in the two orthogonal directions. This allows us to
rewrite (\ref{Top Lagr}) as 
\begin{equation}\label{LT Lagr body}
\bL(g,\dot{g})=\cL^{(l)}(P,\Omega)=\frac{1}{2}\,\langle\Omega,\Omega\rangle
-\frac{1-\alpha}{2}\,\langle\Omega,A\rangle^2-\langle P,A\rangle\;.
\end{equation}
In this case the equations of motion (\ref{Top body}) clearly imply that
the following function is an integral of motion:
\[
C=\langle M,A\rangle\;.
\]
This assures the complete integrability of the Lagrange top.
\vspace{1.5mm}

{\bf Remark 1.} It is easy to see that (\ref{LT inertia}) implies
$\langle M,A\rangle=\alpha\langle\Omega,A\rangle$, which allows us to invert
(\ref{LT inertia}) immediately:
\begin{equation}\label{LT Omega}
\Omega=M+\frac{1-\alpha}{\alpha}\,\langle M,A\rangle\,A\;.
\end{equation}
For futher reference, we rewrite this as
\begin{equation}\label{LT Omega alt}
\Omega=\frac{1}{\alpha}\,M+\frac{1-\alpha}{\alpha}\,[A,[A,M]]\;.
\end{equation}
This, in turn, allows us to reconstruct the motion of the frame $g(t)$ through
the motion of the reduced variables $M(t)$, $P(t)$ (actually only through
$M(t)$). To this end one has to solve the linear differential equation
\[
\dot{g}=g\,\Omega\;.
\]
\vspace{1.5mm}

{\bf Remark 2.}
As almost all known integrable systems, the Lagrange top has a Lax 
representation \cite{RSTS}, \cite{Au}, the original references are \cite{AM}, 
\cite{R}, \cite{RM}. It is straightforward to 
check the following Lax representation 
for (\ref{Top body}), (\ref{LT inertia}) with the matrices from the loop 
algebra $su(2)[\lambda]$:
\begin{equation}\label{Lax repr body}
\dot{L}(\lambda)=[L(\lambda),U(\lambda)]\;,
\end{equation}
where
\begin{equation}\label{Lax matr body}
L(\lambda)=\lambda^2 A+\lambda M+P\;,\qquad U(\lambda)=\lambda A+\Omega\;.
\end{equation}

\subsection{Rest frame formulation}

With the formula (\ref{LT Lagr body}), we can clearly rewrite the Lagrange 
function only in terms of $\omega=\dot{g}g^{-1}$, $a=gAg^{-1}$:
\begin{equation}\label{LT Lagr rest}
\bL(g,\dot{g})=\cL^{(r)}(a,\omega)=
\frac{1}{2}\,\langle\omega,\omega\rangle
-\frac{1-\alpha}{2}\,\langle\omega,a\rangle^2-\langle p,a\rangle\;.
\end{equation}
The possibility to represent $\bL(g,\dot{g})$ through $\omega$, $a$ is 
equivalent to the invariance of $\bL(g,\dot{g})$ under the action
of $G^{(A)}$ on $TG$ induced by right translations on $G$:
\[
\bL(ge^{cA},\dot{g}e^{cA})=\bL(g,\dot{g})\;.
\]
The geometrical meaning of this action is the rotation around $A$ -- the 
symmetry axis of the top. The Euler--Lagrange equations of motion for such
Lagrange functions read:
\begin{equation}\label{Right inv Lagr}
\left\{\begin{array}{l} \dot{m}=[\omega,m]+[a,\nabla_a \cL^{(r)}]\;,\\ \\
\dot{a}=[\omega,a]\;,\end{array}\right.
\end{equation}
where $m=\nabla_\omega \cL^{(r)}$. We calculate for the Lagrange function
(\ref{LT Lagr rest}):
\begin{equation}\label{LT m}
m=\nabla_\omega \cL^{(r)}=\omega-(1-\alpha)\langle \omega,a\rangle a\;,
\end{equation}
\[
\nabla_a \cL^{(r)}=-(1-\alpha)\langle \omega,a\rangle\omega-p\;.
\]
Putting this into (\ref{Right inv Lagr}), we find:
\begin{equation}\label{LT rest frame}
\left\{\begin{array}{l} \dot{m}=[p,a]\\
\dot{a}=[m,a]\end{array}\right.
\end{equation}
which is identical with (\ref{LT rest}). According to Proposition 2.7, this 
system is Hamiltonian with respect to the bracket (\ref{PB right red}), whose
coordinate representation coincides with (\ref{e3 br rest}).
\vspace{1.5mm}

{\bf Remark 1.} It follows from (\ref{LT m}) that $\langle m,a\rangle=\alpha
\langle \omega,a\rangle$, so that (\ref{LT m}) can be easily inverted:
\begin{equation}\label{LT omega}
\omega=m+\frac{1-\alpha}{\alpha}\,\langle m,a\rangle \,a\;.
\end{equation}
Recall that $c=\langle m,a\rangle$ is a Casimir function of the underlying
invariant Poisson bracket (\ref{e3 br rest}). Now the latter formula allows us
to reconstruct the frame evolution from the evolution of the reduced variables
$(m,a)$ via integration of the linear differential equation
\[
\dot{g}=\omega g\;.
\]
\vspace{1.5mm}

{\bf Remark 2.} It turns out to be possible to derive from (\ref{LT rest frame}) 
a closed second order differential equation for $a$. Indeed, take the vector 
product of the second equation in (\ref{LT rest}) in order to obtain
\begin{equation}\label{LT m thru a}
m=a\times\dot{a}+ca\;.
\end{equation}
Substituting this into the first equation in (\ref{LT rest}), we find:
\begin{equation}\label{LT axis}
 a\times\ddot{a}+c\dot{a}=p\times a\;.
\end{equation}
\vspace{1.5mm}

{\bf Remark 3.} The Lax representations for (\ref{LT rest frame}) is, of course,
gauge equivalent to the one for the body frame formulation, but is slightly 
simpler than the latter \cite{R}, \cite{RSTS}. It reads:
\begin{equation}\label{Lax repr rest}
\dot{\ell}(\lambda)=[\ell(\lambda),u(\lambda)]\;,
\end{equation}
with the matrices
\begin{equation}\label{Lax matr rest}
\ell(\lambda)=\lambda^2 a+\lambda m+p\;,\qquad u(\lambda)=\lambda a\;.
\end{equation}
In Sect.6 we indicate how this Lax representation can be derived from the
zero curvature representation of the so called Heisenberg magnetic.

\setcounter{equation}{0}
\section{Discrete time Lagrange top}

We now give (in an {\it ad hoc} manner) the discrete Lagrange function
which is claimed to lead to a suitable discretization of the Lagrange top.
The motivation for the choice of this function comes from the geometry of
curves and will be given in the next section. Unlike the continuous time
case, we start with the rest frame formulation.

\subsection{Rest frame formulation}
Consider
\begin{equation}\label{dLT Lagr rest}
{\bbL}(g_k,g_{k+1})=\Lambda^{(r)}(a_k,w_k)=
-\frac{4\alpha}{\varepsilon}\log{\rm tr}(w_k)-\frac{2(1-\alpha)}{\varepsilon}
\log\Big(1+\langle a_k,w_ka_kw_k^{-1}\rangle\Big)-
\varepsilon\langle p,a_k\rangle\;,
\end{equation}
where $a_k$, $w_k$ are defined as in Sect. 3.2: $w_k=g_{k+1}g_k^{-1}$,
$a_k=g_kAg_k^{-1}$. Notice that $\langle a_k,w_ka_kw_k^{-1}\rangle$ in 
(\ref{dLT Lagr rest}) is nothing but $\langle a_k,a_{k+1}\rangle$.
To see that the function (\ref{dLT Lagr rest}) indeed gives a proper 
discretization of (\ref{LT Lagr rest}), we shall need the following simple 
lemma.

\begin{lemma}
Let $w(\varepsilon)=\ed+\varepsilon\omega+O(\varepsilon^2)\in SU(2)$ be a 
smooth curve, $\omega\in su(2)$. Then
\begin{equation}\label{tr as}
{\rm tr}(w(\epsilon))=2-\frac{\varepsilon^2}{4}\,\langle\omega,\omega\rangle 
+O(\varepsilon^3)\;.
\end{equation}
For an arbitary $a\in su(2)$:
\begin{equation}\label{scalpr as}
\langle a,w(\varepsilon)aw^{-1}(\varepsilon)\rangle=
\langle a,a\rangle-\frac{\varepsilon^2}{2}\Big(\langle a,a\rangle
\langle \omega,\omega\rangle -\langle a,\omega\rangle^2\Big)
+O(\varepsilon^3)\;.
\end{equation}
\end{lemma}
{\bf Proof.} Let $w=\ed+\varepsilon\omega+\varepsilon^2 v+O(\varepsilon^3)$. 
Then from $ww^*=\ed$ we get:
\begin{equation}\label{lemma aux}
v+v^*+\omega\omega^*=0\quad\Rightarrow\quad
v=\frac{1}{2}\,\omega^2+v_1\;,\quad v_1\in su(2)\;.
\end{equation}
Hence
\[
{\rm tr}(v)=\frac{1}{2}\,{\rm tr}(\omega^2)
=-\frac{1}{4}\,\langle\omega,\omega\rangle,
\]
which proves (\ref{tr as}). Similarly, we derive from (\ref{lemma aux}):
\[
waw^*=a+\varepsilon[\omega,a]+\frac{\varepsilon^2}{2}\,[\omega,[\omega,a]]
+\varepsilon^2[v_1,a]+O(\epsilon^3)\;,
\]
which implies (\ref{scalpr as}). \qed
\vspace{1.5mm}

With the help of this lemma we immediately see that, if 
$w=\ed+\varepsilon\omega+O(\varepsilon^2)$, then, up to an additive constant,
\[
\Lambda^{(r)}(a,w)=\varepsilon \cL^{(r)}(a,\omega)+O(\varepsilon^2)\;,
\]
where $\cL^{(r)}(a,\omega)$ is the Lagrange function (\ref{LT Lagr rest})
of the Lagrange top.

\begin{theorem}
The Euler--Lagrange equations of motion for the Lagrange function 
{\rm(\ref{dLT Lagr rest})} are equivalent to the following system:
\begin{equation}\label{dLT rest frame}
\left\{\begin{array}{l} 
m_{k+1}=m_k+\varepsilon[p,a_k]\;,\\ \\
a_{k+1}=a_k+\displaystyle\frac{\varepsilon}{2}\,[m_{k+1},a_k+a_{k+1}]\;.
\end{array}\right.
\end{equation}
The second equation of motion can be uniquely solved for $a_{k+1}$:
\begin{equation}\label{dLT rest frame solved}
a_{k+1}=(\ed+\varepsilon m_{k+1})a_k(\ed+\varepsilon m_{k+1})^{-1}\;.
\end{equation}
The map $(m_k,a_k)\mapsto(m_{k+1},a_{k+1})$ is Poisson with respect to the
bracket {\rm(\ref{e3 br rest})} and has two integrals in involution assuring 
its complete integrability: $\langle m,p\rangle$ and
\begin{equation}\label{dLT rest H}
H_{\varepsilon}(m,a)=\frac{1}{2}\langle m,m\rangle +\langle a,p\rangle
+\frac{\varepsilon}{2}\langle [a,m],p\rangle\;.
\end{equation} 
\end{theorem}
{\bf Proof.} According to Proposition 3.7, the Euler--Lagrange equations of 
motion have the form:
\begin{equation}\label{Right inv dLagr}
\left\{\begin{array}{l} 
w_k^{-1}m_{k+1}w_k=m_k+[a_k,\nabla_a \Lambda^{(r)}(a_k,w_k)]\;,\\ \\
a_{k+1}=w_ka_kw_k^{-1}\;,\end{array}\right.
\end{equation}
where
\begin{equation}\label{dLT m}
m_{k+1}=d_w\Lambda^{(r)}(a_k,w_k)\;.
\end{equation}
To calculate the derivatives of $\Lambda^{(r)}$, we use the following formulas:
\begin{equation}\label{deriv aux1}
d_w{\rm tr}(w_k)=-\frac{1}{2}\,\Im(w_k)\;,\qquad
d_w\langle a_k,w_ka_kw_k^{-1}\rangle=[a_{k+1},a_k]\;.
\end{equation}
\begin{equation}\label{deriv aux2}
\nabla_a\langle a_k,w_ka_kw_k^{-1}\rangle=a_{k+1}+w_k^{-1}a_kw_k\;.
\end{equation}
Indeed, the first one of these expressions follows from:
\[
\langle d_w{\rm tr}(w_k),\eta\rangle=\left.\frac{d}{d\epsilon}{\rm tr}
(e^{\epsilon\eta}w_k)\right|_{\epsilon=0}={\rm tr}(\eta w_k)=
{\rm tr}(\eta\Im(w_k))=-\frac{1}{2}\langle \Im(w_k),\eta\rangle\;.
\]
To prove the second one, proceed similarly:
\[
\langle d_w\langle a_k,w_ka_kw_k^{-1}\rangle,\eta\rangle = 
\left.\frac{d}{d\epsilon}\langle a_k, e^{\epsilon\eta}w_ka_kw_k^{-1}
e^{-\epsilon\eta}\rangle\right|_{\epsilon=0} 
=\langle a_k,[\eta,a_{k+1}]\rangle=
\langle [a_{k+1},a_k],\eta\rangle\;.
\]
Finally, as for the third expression, we have:
\[
\langle \nabla_a\langle a_k,w_ka_kw_k^{-1}\rangle,\eta\rangle =
\left.\frac{d}{d\epsilon}\langle a_k+\epsilon\eta, 
w_k(a_k+\epsilon\eta)w_k^{-1}\rangle\right|_{\epsilon=0} 
 =\langle a_{k+1}+w_k^{-1}a_kw_k,\eta\rangle\;.
\]

With the help of (\ref{deriv aux1}), (\ref{deriv aux2}) we find the following 
expressions:
\begin{equation}\label{dLT m expr}
m_{k+1}=\frac{2\alpha}{\varepsilon}\,\frac{\Im(w_k)}{{\rm tr}(w_k)}-
\frac{2(1-\alpha)}{\varepsilon}\,
\frac{[a_{k+1},a_k]}{1+\langle a_k,a_{k+1}\rangle}\;,
\end{equation}
and
\begin{eqnarray}
w_k^{-1}m_{k+1}w_k-[a_k,\nabla_a\Lambda^{(r)}(a_k,w_k)] & = &
\frac{2\alpha}{\varepsilon}\,\frac{\Im(w_k)}{{\rm tr}(w_k)}-
\frac{2(1-\alpha)}{\varepsilon}\,
\frac{[a_{k+1},a_k]}{1+\langle a_k,a_{k+1}\rangle}+\varepsilon[a_k,p]
\nonumber\\ \nonumber\\
 & = & m_{k+1}+\varepsilon[a_k,p]\;.
\end{eqnarray}
Comparing the latter formula with the first equation of motion in 
(\ref{Right inv dLagr}), we find that it can be rewritten as
\[
m_{k+1}+\varepsilon[a_k,p]=m_k\;,
\]
which is equivalent to the first equation of motion in (\ref{dLT rest frame}).

To derive the second one, rewrite the second equation in (\ref{Right inv dLagr})
as
\begin{eqnarray*}
0=a_{k+1}w_k-w_ka_k & = & \Re(w_k)(a_{k+1}-a_k)+a_{k+1}\Im(w_k)-\Im(w_k)a_k\\
 & = & \frac{1}{2}{\rm tr}(w_k)(a_{k+1}-a_k)+\frac{1}{2}[a_{k+1}+a_k,\Im(w_k)]
\end{eqnarray*}
(we used Lemma C.3 and the equality $\langle a_{k+1}\,,\Im(w_k)\rangle=
\langle a_k\,,\Im(w_k)\rangle$ which follows from the same equation
$a_{k+1}w_k=w_ka_k$ we started with). So, the second equation in (\ref{Right 
inv dLagr}) is equivalent to
\begin{equation}\label{dLT rest aux1}
a_{k+1}-a_k=\left[\,\frac{\Im(w_k)}{{\rm tr}(w_k)}\,,\, a_{k+1}+a_k\right]\;.
\end{equation}
On the other hand, for any two unit vectors $a_k, a_{k+1}$ with 
$a_{k+1}+a_k\neq 0$ we have:
\begin{equation}\label{dLT rest aux2}
a_{k+1}-a_k=-\left[\,\frac{[a_{k+1},a_k]}{1+\langle a_k,a_{k+1}\rangle}\,,
a_{k+1}+a_k\right]\;.
\end{equation}
Comparing (\ref{dLT rest aux1}), (\ref{dLT rest aux2}) with
(\ref{dLT m expr}), we find the second equation of motion in (\ref{dLT rest 
frame}). 

Next, we want to show how the second equation of motion in (\ref{dLT rest 
frame}) can be solved for $a_{k+1}$. This equation implies 
$\langle a_{k+1},m_{k+1}\rangle=\langle a_k,m_{k+1}\rangle$, so that,
according to Lemma C.3, it can be rewritten as
\[
a_{k+1}+\varepsilon a_{k+1}m_{k+1}=a_k+\varepsilon m_{k+1}a_k\;,
\]
which is clearly equivalent to (\ref{dLT rest frame solved}).

The Poisson properties of the map (\ref{dLT rest frame}) are assured 
by Proposition 3.7. 

It remains to demonstrate that the function (\ref{dLT rest H}) is indeed an
integral of motion. This is done by the following derivation:
\begin{eqnarray*}
H_{\varepsilon}(m_{k+1},a_{k+1}) & = & 
\frac{1}{2}\langle m_{k+1},m_{k+1}\rangle +\langle a_{k+1}+\frac{\varepsilon}
{2}\,[a_{k+1},m_{k+1}]\,,p\,\rangle \\
& = & \frac{1}{2}\langle m_{k+1},m_{k+1}\rangle +\langle a_k-\frac{\varepsilon}
{2}\,[a_k,m_{k+1}]\,,p\,\rangle\\
& = & \frac{1}{2}\langle m_{k+1},m_{k+1}-\varepsilon [p,a_k]\,\rangle 
+\langle a_k,p\,\rangle\\
& = & \frac{1}{2}\langle m_k+\varepsilon [p,a_k], m_k\rangle 
+\langle a_k,p\,\rangle \;=\; H_{\varepsilon}(m_k,a_k)\;.\\ 
\end{eqnarray*}
The theorem is proved. \qed
\vspace{1.5mm}

{\bf Remark.} The equations of motion (\ref{dLT rest frame}), being written
entirely in terms of elements of the Lie algebra $su(2)$, are clearly
equivalent to the equations of motion (\ref{dLT rest}), which are written
in terms of vectors from ${\Bbb R}^3$. The situation with (\ref{dLT rest
frame solved}) is slightly different. Indeed, it corresponds to the following 
formula in ${\Bbb R}^3$:
\[
a_{k+1}=Q_{k+1}a_k=
\frac{\ed+\varepsilon\bm_{k+1}/2}{\ed-\varepsilon\bm_{k+1}/2}\,a_k\;,
\]
where the orthogonal matrix $Q_{k+1}\in SO(3)$ is constructed out of the
skew--symmetric matrix $\bm_{k+1}\in so(3)$ which corresponds to the vector
$m_{k+1}\in{\Bbb R}^3$ according to the following rule:
\[
m=(m_1,m_2,m_3)^T\in{\Bbb R}^3 \;\;\leftrightarrow\;\;
\bm=\left(\begin{array}{ccc} 0 & -m_3 & m_2 \\ m_3 & 0 & -m_1 \\
-m_2 & m_1 & 0 \end{array}\right)\in so(3)\;.
\]
\vspace{2mm}

Just as in the continuous time case, it is possible to derive a closed
second order difference equation for the motion of the body axis $a_k$.
\begin{proposition} The sequence of $a_k$ satisfies the following equation:
\begin{equation}\label{dLT eq for axis}
a_k\times\left(\frac{2\,a_{k+1}}{1+\langle a_k,a_{k+1}\rangle}
+\frac{2\,a_{k-1}}{1+\langle a_{k-1},a_k\rangle}\right)+
\varepsilon c \left(\frac{a_{k+1}+a_k}{1+\langle a_k,a_{k+1}\rangle}-
\frac{a_k+a_{k-1}}{1+\langle a_{k-1},a_k\rangle}\right)=
\varepsilon^2p \times a_k\;,
\end{equation}
where $c=\langle m_k,a_k\rangle$ is an integral of motion.
\end{proposition}
{\bf Proof.} Take a vector product of the second equation of motion in
(\ref{dLT rest}) by $a_{k+1}+a_k$. Taking into account that 
$\langle m_{k+1},a_{k+1}\rangle=\langle m_{k+1},a_k\rangle=c$, we find:
\[
2a_k\times a_{k+1}=\varepsilon m_{k+1}(1+\langle a_k,a_{k+1}\rangle)-
\varepsilon c(a_{k+1}+a_k)\;,
\]
or
\begin{equation}\label{dLT rest aux3}
m_{k+1}=
\frac{2}{\varepsilon}\,\frac{a_k\times a_{k+1}}{1+\langle a_k,a_{k+1}\rangle}
+c\,\frac{a_{k+1}+a_k}{1+\langle a_k,a_{k+1}\rangle}\;.
\end{equation}
Plugging this into the first equation of motion in (\ref{dLT rest}), we arrive
at (\ref{dLT eq for axis}). \qed
\vspace{1.5mm}

Further, we demonstrate how to reconstruct the ``angular velocity'' $w_k$
(and therefore the motion of the frame $g_k$) from the evolution of the
reduced variables $(a_k,m_k)$. 
\begin{proposition}
The discrete time evolution of the frame $g_k$ can be determined from the 
linear difference equation
\begin{equation}\label{ev g}
g_{k+1}=w_kg_k\;,
\end{equation}
where $w_k$ are given by
\begin{equation}\label{ev w}
w_k=\frac{{\rm tr}(w_k)}{2}(\ed+\varepsilon\xi_k)\;,
\end{equation}
where
\begin{equation}\label{ev xi}
\xi_k=m_{k+1}+c\,\frac{1-\alpha}{\alpha}\,\frac{a_{k+1}+a_k}
{1+\langle a_k,a_{k+1}\rangle}=
\frac{2}{\varepsilon }\,\frac{a_k\times a_{k+1}}{1+\langle a_k,a_{k+1}\rangle}
+\frac{c}{\alpha}\,\frac{a_{k+1}+a_k}{1+\langle a_k,a_{k+1}\rangle} 
\;,
\end{equation} 
and
\begin{equation}\label{tr w}
{\rm tr}(w_k)=\frac{2}{\sqrt{1+\displaystyle
\frac{\varepsilon^2}{4}\langle\xi_k,\xi_k\rangle}}=
\sqrt{2\,\displaystyle\frac{1+\langle a_k,a_{k+1}\rangle}
{1+\varepsilon^2 c^2/4\alpha^2}}\;.
\end{equation}
\end{proposition}
{\bf Proof.} We combine (\ref{dLT m expr})
with (\ref{dLT rest aux3}) in order to derive the formula
\[
2\,\frac{\Im(w_k)}{{\rm tr}(w_k)}=\varepsilon\xi_k
\]
with the expressions for $\xi_k$ given in (\ref{ev xi}). Now the reference
to Lemma C.2 finishes the proof. \qed
\vspace{1.5mm}

Finally, we give a Lax representation for the map (\ref{dLT rest frame}).
\begin{theorem}
The map {\rm(\ref{dLT rest frame})} has the following Lax representation:
\begin{equation}\label{dLT rest Lax}
\ell_{k+1}(\lambda)=u_k^{-1}(\lambda)\ell_k(\lambda)u_k(\lambda)\;,
\end{equation}
with the matrices
\begin{equation}\label{dLT rest Lax matrices}
\ell_k(\lambda)
=\lambda^2\left(a_k+\frac{\varepsilon}{2}\,[a_k,m_k]+\frac{\varepsilon^2}
{4}\,p\right)+\lambda m_k+p\;,\qquad u_k(\lambda)=\ed+\varepsilon\lambda a_k\;.
\end{equation}
\end{theorem}
{\bf Proof} -- a direct verification. \qed

In the next section we present a derivation of this Lax representation from
the one for the so called lattice Heisenberg magnetic.

\subsection{Moving frame formulation}
Note that the discrete Lagrange function (\ref{dLT Lagr rest}) may be also
expressed in terms of $P_k=g_k^{-1}pg_k$, $W_k=g_k^{-1}g_{k+1}$:
\begin{equation}\label{dLT Lagr body}
{\bbL}(g_k,g_{k+1})=\Lambda^{(l)}(P_k,W_k)=
-\frac{4\alpha}{\varepsilon}\log{\rm tr}(W_k)-\frac{2(1-\alpha)}{\varepsilon}
\log\Big(1+\langle A,W_k^{-1}AW_k\rangle\Big)-
\varepsilon\langle P_k,A\rangle\;.
\end{equation}
Since $W_k=\ed+\varepsilon\Omega+O(\varepsilon^2)$, we can apply Lemma 5.1
to see that
\[
\Lambda^{(l)}(P_k,W_k)=\varepsilon\cL^{(l)}(P,\Omega)+O(\varepsilon^2)\;,
\]
where $\cL^{(l)}(P,\Omega)$ is the Lagrange function (\ref{LT Lagr body})
of the continuous time Lagrange top. Now, one can derive all results concerning
the discrete time Lagrange top in the body frame from the ones in the rest
frame by performing the change of frames so that
\[
M_k=g_k^{-1}m_kg_k\;,\qquad P_k=g_k^{-1}pg_k\;,\qquad A=g_k^{-1}a_kg_k\;.
\]
\begin{theorem}
The Euler--Lagrange equations for the Lagrange function {\rm(\ref{dLT Lagr body})}
are equivalent to the following system:
\begin{equation}\label{dLT body}
\left\{\begin{array}{l}
M_{k+1}=W_k^{-1}\Big(M_k+\varepsilon[P_k,A]\,\Big)W_k\;,\\ \\
P_{k+1}=W_k^{-1}P_kW_k\;,
\end{array}\right.
\end{equation}
where the ``angular velocity'' $W_k$ is determined by the ``angular momentum''
$M_{k+1}$ via the following formula and Lemma C.2:
\begin{eqnarray}\label{dLT W thru M}
2\,\frac{\Im(W_k)}{{\rm tr}(W_k)} & = & \frac{\varepsilon}{\alpha}\,M_{k+1}+
\frac{2(1-\alpha)}{\alpha}\,\frac
{\Big[A,\,(\ed+\varepsilon M_{k+1})^{-1}A(\ed+\varepsilon M_{k+1})\Big]}
{1+\Big\langle A,\,(\ed+\varepsilon M_{k+1})^{-1}A(\ed+\varepsilon M_{k+1})\Big
\rangle}\\
 & = & \varepsilon\left(\frac{1}{\alpha}\,M_{k+1}+\frac{1-\alpha}{\alpha}\,
[A,[A,M_{k+1}]]\right)+O(\varepsilon^2)\;.\nonumber
\end{eqnarray}
The map {\rm(\ref{dLT body}), (\ref{dLT W thru M})} is Poisson with respect
to the Poisson bracket {\rm(\ref{e3 br body})} and has two integrals in 
involution assuring its complete integrability: $\langle M,A\rangle$ and
\begin{equation}\label{dLT body H}
H_{\varepsilon}(M,P)=\frac{1}{2}\langle M,M\rangle +\langle P,A\rangle
+\frac{\varepsilon}{2}\langle [M,P],A\rangle\;.
\end{equation} 
\end{theorem}

{\bf Remark.} It might be preferable to express $W_k$ through $(M_k,P_k)$
rather than through $M_{k+1}$ (in particular, this is necessary in order to
demonstrate that the map $(M_k,P_k)\mapsto(M_{k+1},P_{k+1})$ is well defined).
The corresponding expression reads:
\begin{equation}\label{dLT WMW expr}
2\,\frac{\Im(W_k)}{{\rm tr}(W_k)}=
\frac{\varepsilon}{\alpha}\,(M_k+\varepsilon[P_k,A])-
\frac{2(1-\alpha)}{\alpha}\,\frac{[A,\,W_kAW_k^{-1}]}{1+\langle
A,\,W_kAW_k^{-1}\rangle}\;,
\end{equation}
\begin{equation}\label{dLT body aux6}
W_kAW_k^{-1}=(\ed+\varepsilon M_k+\varepsilon^2[P_k,A])\,A\,
(\ed+\varepsilon M_k+\varepsilon^2 [P_k,A])^{-1}\;.
\end{equation}
We see that the resulting formula is similar to (\ref{dLT W thru M}), but
its right--hand side depends not only on $M_k$ but also on $P_k$ (though
this latter dependence appears only in $O(\varepsilon^2)$ terms). According to 
Lemma C.2, both versions allow for the reconstruction of the evolution
of the frame $g_k$ from the evolution of the reduced variables $(M_k,P_k)$,
anyway.
\vspace{2mm}

We close this section with a Lax representation for the map (\ref{dLT body}),
(\ref{dLT W thru M}).
\begin{theorem}
The map {\rm(\ref{dLT body}), (\ref{dLT W thru M})} has the following Lax 
representation:
\begin{equation}\label{dLT body Lax}
L_{k+1}(\lambda)=U_k^{-1}(\lambda)L_k(\lambda)U_k(\lambda)\;,
\end{equation}
with the matrices
\begin{equation}\label{dLT body Lax matrices}
L_k(\lambda)
=\lambda^2\left(A+\frac{\varepsilon}{2}\,[A,M_k]+\frac{\varepsilon^2}
{4}\,P_k\right)+\lambda M_k+P_k\;,\qquad 
U_k(\lambda)=(\ed+\varepsilon\lambda A)W_k\;.
\end{equation}
\end{theorem}
{\bf Proof} -- a direct verification. \qed

\setcounter{equation}{0}
\section{Motivation: Lagrange top and elastic curves}

The Lagrange function (\ref{dLT Lagr rest}) was found using an analogy
between the Lagrange top and the elastic curves as a heuristic tool. The
present section is devoted to an exposition of the corresponding
interrelations.

Let $\gamma:\,[0,l]\mapsto {\Bbb R}^3$ be a smooth curve parametrized by
the arclength $x\in [0,l]$. Defining the tangent vector $T:\,[0,l]\mapsto 
{\Bbb R}^3$ as $T(x)=\gamma'(x)$, the characteristic property of the arclength
parametrization may be expressed as 
\begin{equation}\label{|T|}
|T(x)| =1\;,
\end{equation}
where $|\cdot|$ stands for the euclidean norm. The {\it curvature} of the 
curve $\gamma$ is defined as 
\begin{equation}\label{curvature}
k(x)=|T'(x)|\,.
\end{equation}
\begin{definition} {\rm \cite{L}, \cite{LS}}.
A {\em classical elastic curve} (Bernoulli's elastica) is a curve delivering 
an extremum to the functional
\begin{equation}\label{class elastic funct}
\int_0^l k^2(x)dx\;,
\end{equation}
the admissible variations of the curve are those preserving $\gamma(0)$
and $\gamma(l)$, more precisely, those preserving $\gamma(l)-\gamma(0)=
\int_0^l T(x)dx$. 
\end{definition}
Introducing the Lagrange multipliers $p\in{\Bbb R}^3$
corresponding to this constraint, we come to the functional
\begin{equation}\label{class elastic funct mech}
\int_0^l \Big(|T'(x)|^2-2\,\langle p,T(x)\rangle\Big) dx\;.
\end{equation}
Identifying the arclength parameter $x$ with the time $t$,
this functional becomes (twice) the action functional for the {\it spherical
pendulum}. So, classical elasticae are in a one--to--one correspondence with
the motions of the spherical pendulum.

A generalization of these notions to {\it elastic rods} (which physically means
that they can be twisted) requires the curves to be framed, i.e. to carry an 
orthonormal frame $\Phi(x)=(T(x),N(x),$ $B(x))$
in each point. In other words, a {\it framed curve} is a map
$\Phi:\,[0,l]\mapsto \{{\rm frames}\}$. The curve itself is then defined
by integration: $\gamma(x)=\int_0^x T(y)dy$. The following quantities are
attributes of a framed curve: the {\it geodesic curvature}
\begin{equation}\label{geod curvature}
k_1(x)=\langle T'(x),N(x)\rangle\;,
\end{equation} 
the {\it normal curvature}
\begin{equation}\label{normal curvature}
k_2(x)=\langle T'(x), B(x)\rangle\;,
\end{equation}
and the {\it torsion}
\begin{equation}\label{torsion}
\tau(x)=\langle N'(x),B(x)\rangle\;.
\end{equation}
Obviously, one has: $k^2(x)=k_1^2(x)+k_2^2(x)$. 
\begin{definition} {\rm \cite{L}, \cite{LS}}.
An {\em elastic rod} (Kirchhoff's elastica) is a framed 
curve delivering an extremum to the functional
\begin{equation}\label{elastic funct}
\int_0^l \Big(k^2(x)+\alpha \tau^2(x)\Big)dx
\end{equation}
with some $\alpha\neq 0$. The admissible variations of the curve preserve
$\Phi(0)$, $\Phi(l)$, and $\gamma(l)-\gamma(0)=\int_0^l T(x)dx$.
\end{definition}
The first term in (\ref{elastic funct}) corresponds to the bending energy,
the second one corresponds to the twist energy.

We shall identify ${\Bbb R}^3$ with $su(2)$, as described in Appendix C, and 
the frames with elements of $\Phi\in SU(2)$, according to the following 
prescription:
\begin{equation}\label{frame in SU(2)}
T=\Phi^{-1}\be_3\Phi\;, \quad N=\Phi^{-1}\be_1\Phi\;, \quad
B=\Phi^{-1}\be_2\Phi\;.
\end{equation}
Then, denoting
\begin{equation}\label{omegas}
\Omega=-\Phi'\Phi^{-1}\;,\qquad \omega=-\Phi^{-1}\Phi'\;,
\end{equation}
we find:
\begin{equation}\label{curvatures in top}
k_1=\langle \omega,B\rangle=\langle \Omega,\be_2\rangle\,=\Omega_2,\qquad
k_2=-\langle \omega,N\rangle=-\langle \Omega,\be_1\rangle\,=-\Omega_1\;,
\end{equation}
\begin{equation}\label{torsion in top}
\tau=\langle \omega,T\rangle=\langle \Omega,\be_3\rangle=\Omega_3\;.
\end{equation}
So, the variational problem for elastic rods may be formulated as follows:
find $\Phi(x)\,:\,[0,l]\mapsto SU(2)$ delivering an extremum of the functional
\begin{equation}\label{elastic funct in G}
\int_0^l \Big(\Omega_1^2(x)+\Omega_2^2(x)+\alpha \Omega_3^2(x)-2\,
\langle p,T(x)\rangle\Big)dx\;,
\end{equation}
where $p$ is an ($x$--independent) Lagrange multiplier coming from the
condition of fixed $\gamma(l)-\gamma(0)=\int_0^l T(x)dx$. Identifying the 
arclength parameter $x$ with the time $t$, $\Phi(x)=g^{-1}(t)$, so that 
$\Omega(x)=-\Phi'(x)\Phi^{-1}(x)=g^{-1}(t)\dot{g}(t)=\Omega(t)$, and 
$T(x)=\Phi^{-1}(x)\be_3\Phi(x)=g(t)\be_3g^{-1}(t)=a(t)$, we see that
the functional (\ref{elastic funct in G}) coincides with (twice) the action
functional for the Lagrange top. This proves the 
\begin{proposition} {\em (Kirchhoff's kinetic analogy, \cite{L})}. 
The frames of arclength parametrized elastic rods are in a one--to--one 
correspondence with the motions of the Lagrange top.
\end{proposition}

Actually, we use another characterization of the elastic rods.
From the Euler--Lagrange equations it follows:
\begin{proposition} The torsion $\tau$ along the extremals of the functional
{\rm(\ref{elastic funct in G})} is constant, and the tangent vector $T(x)$ 
satisfies the following second--order differential equation:
\begin{equation}\label{second order}
T\times T''+cT'=p\times T\;,
\end{equation} 
where $c=\alpha\tau$.
Conversely, each solution $T(x)$ of {\rm (\ref{second order})} corresponds to
a curve $\gamma(x)$, which, being equipped with a frame with constant
torsion $\tau$, delivers an extremum to the functional {\rm(\ref{elastic funct 
in G})} with $\alpha=c/\tau$.
\end{proposition}

Equation (\ref{second order}) is (\ref{LT axis}) in new 
notations. The latter differential equation allows the following 
interpretation. Consider the so called {\it Heisenberg flow}. 
It is defined by the differential equation
\begin{equation}\label{Heis}
T_{\bf t}=T\times T''\;,
\end{equation} 
and describes the evolution of a curve in the binormal direction with the
velocity equal to the curvature. Here the ``time'' {\bf t} has nothing in 
common with the time $t$ of the Lagrange top, which is, remember, identified 
with $x$. It is easy to see that the flow on curves defined by the vector 
field $T_x=T'$ (a reparametrization of a curve) commutes with the Heisenberg 
flow (\ref{Heis}). Using this fact, we can integrate (\ref{Heis}) once in order 
to find
\begin{equation}\label{LIE}
\gamma_{\bf t}=\gamma'\times\gamma''=T\times T'\;.
\end{equation} 
(The reparametrization flow, once integrated, takes the form $\gamma_x=
\gamma'=T$).
Now we can formulate the following fundamental statement. 
\begin{theorem} {\rm \cite{Ha}, \cite{LS}}.
Let $\Phi\,:\,[0,l]\mapsto SU(2)$ be the frame of an elastic rod, and 
$\gamma\,:\,[0,l]\mapsto su(2)$ the corresponding curve with the tangent
vector $T=\gamma'\,:[0,l]\mapsto su(2)$. Then the evolution of $\gamma$ 
under the Heisenberg flow {\rm(\ref{LIE})} is a rigid screw--motion, and the 
evolution of $T$ under the Heisenberg flow {\rm(\ref{Heis})} is a rigid 
rotation. Conversely, if the evolution of $T$ is a rigid rotation, then $T$ 
can be lifted to a frame $\Phi$ of an elastic rod.
\end{theorem}

The first statement of the theorem follows from (\ref{second order}). The
left--hand side of (\ref{second order}) can be interpreted as the vector field 
on curves, corresponding to a linear combination of the Heisenberg flow and 
the reparametrization:
\[
T_{\bf t}+cT_x=p\times T\;. 
\] 
Integrated once, this equation yields a rigid screw motion for the curve 
$\gamma$:
\[
\gamma_{\bf t}+c\gamma_x=p\times \gamma+q\;,
\] 
where $q\in su(2)$ is a fixed vector. The converse statement follows from
Proposition 6.4.

By the way, this theorem allows to find a Lax representation for the
equation (\ref{second order}), and therefore for the Lagrange top, starting
from the well--known Lax representation for the Heisenberg flow.
\begin{proposition}
The equation {\rm(\ref{second order})} is equivalent to the Lax equation
\begin{equation}\label{Heis stat Lax}
\ell_x(\lambda)=[\ell(\lambda),u(\lambda)]
\end{equation}
with the matrices
\begin{equation}\label{Heis stat Lax matr}
\ell(\lambda)=\lambda^2 T+\lambda(T\times T_x+cT)+p\;, 
\qquad u(\lambda)=\lambda T\;.
\end{equation}
\end{proposition}
{\bf Proof.} Indeed, the Heisenberg flow (\ref{Heis}) is equivalent to the 
following matrix equation (``zero curvature representation'', \cite{FT}):
\[
u_{\bf t}-v_x+[v,u]=0\;,
\]
where $u,v\in su(2)[\lambda]$ are the following matrices:
\[
u=\lambda T\;,\qquad v=\lambda^2 T+\lambda T\times T_x\;.
\]
Now it is easy to derive that the equation (\ref{second order}), rewritten
as
\[
T_{\bf t}+cT_x=[p,T]\;,
\]
is equivalent to (\ref{Heis stat Lax}) with $\ell=v+cu+p$. \qed

Remembering that in the Kirchhoff's kinetic analogy $x$ is identified with
$t$, $T$ is identified with $a$, and recalling the formula (\ref{LT m thru a}),
we recover the Lax representation of the Lagrange top in the rest frame
given in (\ref{Lax repr rest}), (\ref{Lax matr rest}).
\vspace{2mm}

Theorem 6.5 is also a departure point for discretizing elastic curves and,
therefore, the Lagrange top \cite{B}. A {\it discrete arc--length parametrized
curve} is a sequence $\gamma\,:\,{\Bbb Z}\mapsto {\Bbb R}^3$ with the
property $|T_k|=1$, where $T_k=\gamma_k-\gamma_{k-1}$. Correspondingly, 
{\it discrete framed curves} are the sequences of orthonormal frames $\Phi_k$, 
such that $T_k=\Phi_k^{-1}\be_3\Phi_k$ \footnote{Note that the frames $\Phi_k$, 
as well as the tangent vectors $T_k$, are attached to the edges $[\gamma_{k-1},
\gamma_k]$ of the discrete curve $\gamma$}. As before, we identify ${\Bbb R}^3$ 
with $su(2)$, and the space of orthonormal frames with $SU(2)$. The curve 
$\gamma$ can be reconstructed by applying the summation operation to the 
sequence $T$. 

A discretization of the Heisenberg flow is well known \cite{Skl}, \cite{FT},
see also \cite{DS} for geometric interpretation of discrete flow. It reads:
\begin{equation}\label{dHeis}
(T_k)_{\bf t}=\frac{2\,T_k\times T_{k+1}}{1+\langle T_k,T_{k+1}\rangle}-
\frac{2\,T_{k-1}\times T_k}{1+\langle T_{k-1},T_k\rangle}\;.
\end{equation}
A commuting flow approximating $T_x=T'$ is given by:
\begin{equation}\label{dTx}
(T_k)_{x}=\frac{T_k+T_{k+1}}{1+\langle T_k,T_{k+1}\rangle}-
\frac{T_{k-1}+T_k}{1+\langle T_{k-1},T_k\rangle}\;.
\end{equation}  
Once ``integrated'', this gives the flows on $\gamma_k$:
\begin{equation}\label{dLie}
(\gamma_k)_{x}=\frac{T_k+T_{k+1}}{1+\langle T_k,T_{k+1}\rangle}\;,
\qquad
(\gamma_k)_{\bf t}=\frac{2\,T_k\times T_{k+1}}{1+\langle T_k,T_{k+1}\rangle}\;.
\end{equation} 

Now we accept the following discrete version of Theorem 6.5 as a definition
of discrete elastic rods.
\begin{definition} A {\em discrete elastic rod} is a framed curve for which
the evolution of $\gamma_k$ under a linear combination of flows 
$(\gamma_k)_{\bf t}+c(\gamma_k)_x$ with some $c$ is a rigid
skrew--motion, so that the evolution of $T_k$ under the flow 
$(T_k)_{\bf t}+c(T_k)_x$ is a rigid rotation. 
\end{definition}
In other words, 
the sequence $T_k$ satisfies the following second order difference equation:
\begin{equation}\label{dsecond}
T_k\times\left(\frac{2\,T_{k+1}}{1+\langle T_k,T_{k+1}\rangle}+
\frac{2\,T_{k-1}}{1+\langle T_{k-1},T_k\rangle}\right)
+c \left(\frac{T_{k+1}+T_k}{1+\langle T_k,T_{k+1}\rangle}-
\frac{T_k+T_{k-1}}{1+\langle T_{k-1},T_k\rangle}\right)
=p\times T_k\;.
\end{equation}
(This is the equation (\ref{dLT eq for axis}) with $\varepsilon=1$ in new 
notations). We can immediately find the Lax representation for the difference
equation (\ref{dsecond}).
\begin{proposition}
The equation {\rm(\ref{dsecond})} is equivalent to the Lax equation
\begin{equation}\label{dHeis stat Lax}
\ell_{k+1}(\lambda)=u_k^{-1}(\lambda)\ell_k(\lambda)u_k(\lambda)
\end{equation}
with the matrices
\begin{equation}\label{dHeis stat L}
(1+\lambda^2/4)\,\ell_k(\lambda)=(\lambda^2+c\lambda)\,\frac{T_k+T_{k-1}}
{1+\langle T_k,T_{k-1}\rangle}+
(2\lambda-c\lambda^2/2)\,\frac{T_{k-1}\times T_k}
{1+\langle T_k,T_{k-1}\rangle}+(1+\lambda^2/4)\,p\;, 
\end{equation}
\begin{equation}\label{dHeis stat U}
u_k(\lambda)=\ed+\lambda T_k\;.
\end{equation}
\end{proposition}
{\bf Proof.} It is well known (see \cite{FT}) that the flows (\ref{dHeis}),
(\ref{dTx}) allow the following ``discrete zero curvature representations'':
\[
(u_k)_{\bf t}=u_kv_{k+1}^{(1)}-v_k^{(1)}u_k\;,\qquad
(u_k)_x=u_kv_{k+1}^{(0)}-v_k^{(0)}u_k\;,
\] 
respectively, with the matrices $u_k$ as in (\ref{dHeis stat U}) and
\[
(1+\lambda^2/4)\,v_k^{(1)}=\lambda^2\,\frac{T_k+T_{k-1}}
{1+\langle T_k,T_{k-1}\rangle}+2\lambda\,\frac{T_{k-1}\times T_k}
{1+\langle T_k,T_{k-1}\rangle}\;,
\]
\[
(1+\lambda^2/4)\,v_k^{(0)}=\lambda\,\frac{T_k+T_{k-1}}
{1+\langle T_k,T_{k-1}\rangle}-\frac{\lambda^2}{2}\,\frac{T_{k-1}\times T_k}
{1+\langle T_k,T_{k-1}\rangle}\;.
\]
Now it is easy to see that the equation (\ref{dsecond}), rewritten as
\[
(T_k)_{\bf t}+c(T_k)_x=[p,T_k]\;,
\]
is equivalent to $u_k\ell_{k+1}=\ell_ku_k$ with
\[
\ell_k=v_k^{(1)}+cv_k^{(0)}+p\;,
\]
which coincides with (\ref{dHeis stat L}). \qed

To establish a link with the discrete time Lagrange top, recall that the
formula (\ref{dLT rest aux3}) in our new notations reads:
\[
m_k=2\,\frac{T_{k-1}\times T_k}{1+\langle T_k,T_{k-1}\rangle}+
c\,\frac{T_k+T_{k-1}}{1+\langle T_k,T_{k-1}\rangle}\;,
\]
which implies also
\[
T_k+\frac{1}{2}\,T_k\times m_k=\frac{T_k+T_{k-1}}{1+\langle T_k,T_{k-1}\rangle}
-\frac{c}{2}\,\frac{T_{k-1}\times T_k}{1+\langle T_k,T_{k-1}\rangle}\;.
\]
Hence we can write:
\[
(1+\lambda^2/4)\,\ell_k=
\lambda^2\Big(T_k+\frac{1}{2}\,T_k\times m_k+\frac{1}{4}\,p\Big)
+\lambda m_k+p\;,
\] 
which coincides with (\ref{dLT rest Lax matrices}) up to a nonessential
constant factor.
\vspace{2mm}

It remains to find a variational problem generating the equations 
of motion (\ref{dsecond}). But the calculations of Sect. 5 show that this task 
is solved by the functional (\ref{dLT Lagr rest}).
This gives the following alternative definition of discrete elastic rods. 
\begin{definition} A {\em discrete  elastic rod} is a discrete framed curve
given by a finite sequence $\Phi_1$,...,$\Phi_N$ $\in su(2)$ delivering an
extremum to the functional
\begin{equation}\label{dfunctional}
\sum_{k=1}^{N-1}\Big(-4\alpha\,\log{\rm tr}(\Phi_k^{-1}\Phi_{k+1})-2(1-\alpha)
\log(1+\langle T_k,T_{k+1}\rangle)\Big)-\sum_{k=1}^N\langle p,T_k\rangle
\end{equation}
with some $\alpha\neq 0$. The admissible variations of the curve preserve
$\Phi_1$, $\Phi_N$, and $\gamma_N-\gamma_0=\sum_{k=1}^N\Phi_k^{-1}\be_3\Phi_k$.
\end{definition}

The equivalence of Definitions 6.7 and 6.9 is the basic new result of this
section. It is a geometric counterpart and a motivation for the considerations 
of Sect.5.

We want to close this section by giving discretizations of geometrical notions 
like curvature and torsion. Notice that the functional (\ref{dfunctional}) 
naturally splits into two parts,  one independent on $\alpha$ and one 
proportional to $\alpha$. Accordingly, we declare 
\begin{equation}\label{discr bend en}
-2\,\sum_k\log(1+\langle T_k,T_{k+1}\rangle)=2\,\sum_k\log\left(1+\frac{1}{4}
k_{k}^2\right)+{\rm const}
\end{equation}
as a discretization of the ``bending energy'' $\frac{1}{2}\int_0^l k^2(x)dx$,
and
\begin{equation}\label{discr tw en}
\sum_k\Big(-4\log\,{\rm tr}(\Phi_{k+1}\Phi_k^{-1})+2
\log(1+\langle T_k,T_{k+1}\rangle)\Big)=2\,\sum_k\log\left(1+\frac{1}{4}
\tau_{k}^2\right)+{\rm const}
\end{equation}
as a discretization of the ``twist energy'' $\frac{1}{2}\int_0^l\tau^2(x)dx$.
Here we define the ``discrete curvature'' $k_{k}$ at the vertex $\gamma_k$ by
\[
1+\frac{1}{4}k_{k}^2=\frac{2}{1+\langle T_k,T_{k+1}\rangle}\;\;
\Leftarrow\;\; 
k_{k}=2\,{\rm tan}(\varphi_k/2)\;,
\]
where $\varphi_k$ is the angle between the vectors $T_k$ and $T_{k+1}$.
Notice that the $k_k$ depends not on the whole frame, but on the tangent 
vectots $T_k$ only, so that it makes sense also for non--framed curves. 
The ``discrete torsion'' $\tau_{k}$ at the vertex $\gamma_k$ is defined by
\[
1+\frac{1}{4}\tau_{k}^2=\frac{2(1+\langle T_k,T_{k+1}\rangle)}
{({\rm tr}(\Phi_{k+1}\Phi_k^{-1}))^2}\;\;\Leftarrow\;\;
\tau_k=-2\left\langle\frac{\Im(\Phi_{k+1}\Phi_k^{-1})}
{{\rm tr}(\Phi_{k+1}\Phi_k^{-1})},\,\be_3\right\rangle\;.
\]
The last formula will be commented on immediately.
Let us demonstrate that, in a complete analogy with the continuous case,
the discrete torsion is constant along the extremals of the functional
(\ref{dfunctional}). Denoting for a moment
\[
\Phi_{k+1}\Phi_k^{-1}=\left(\begin{array}{cc} a & b \\ -\bar{b} & \bar{a}
\end{array}\right)\in SU(2)\;,
\]
we find:
\[
1+\langle T_k,T_{k+1}\rangle=1-2\,{\rm tr}(\Phi_{k+1}\Phi_k^{-1}\be_3
\Phi_k\Phi_{k+1}^{-1}\be_3)=2|a|^2\;,
\]
and also
\[
\Re(a)=\frac{1}{2}\,{\rm tr}(\Phi_{k+1}\Phi_k^{-1})\;,\qquad
\Im(a)={\rm tr}(\Phi_{k+1}\Phi_k^{-1}\be_3)=-\frac{1}{2}\langle
\Im(\Phi_{k+1}\Phi_k^{-1}),\,\be_3\rangle\;,
\]
so that 
\begin{equation}\label{dtorsion}
\tau_{k}=2\,\frac{\Re(a)}{\Im(a)}=
-2\left\langle\frac{\Im(\Phi_{k+1}\Phi_k^{-1})}{{\rm tr}(\Phi_{k+1}\Phi_k^{-1})},
\,\be_3\right\rangle=
-2\left\langle\frac{\Im(\Phi_k^{-1}\Phi_{k+1})}{{\rm tr}(\Phi_k^{-1}\Phi_{k+1})},
\,T_{k+1}\right\rangle\;.
\end{equation}
Comparing this with (\ref{dLT m expr}) (remember, we set $\varepsilon=1$ and
identified $a_k$ with $T_k$ and $w_k$ with $\Phi_{k+1}^{-1}\Phi_k$), we see
that
\[
\tau_{k}=c/\alpha\;,
\]
where $c=\langle m_{k+1}, T_{k+1}\rangle$ is an integral of motion of the
Euler--Lagrange equations (a Casimir function of the $e(3)$ Lie--Poisson 
bracket). This corresponds literally to the continuous case.
\vspace{2mm}

{\bf Remark.} The case $\alpha=0$ corresponds to {\em discrete elastic curves}
$\gamma:{\Bbb Z}\mapsto {\Bbb R}^3$. The tangent vectors $T:{\Bbb Z}\mapsto
S^2$, $T_k=\gamma_k-\gamma_{k-1}$, define a trajectory of the {\em discrete 
time spherical pendulum}. Its Lagrange function is obtained, as in the 
continuous time case, from the bending energy (\ref{discr bend en}), upon 
introducing the Lagrange multiplier $p$. Notice that the Lagrange function of 
the discrete time spherical pendulum is defined on $S^2\times S^2$.

\section{Visualisation}
After the theory has been developed, it is tempting to look at the spinning
of the discrete time Lagrange top. Fortunately, in the computer era, a discrete 
time top is even simpler to simulate than a classical one. Indeed, as it is
shown in Theorem 5.2, the Poisson map $(m_k,a_k)\mapsto(m_{k+1},a_{k+1})$ is
well defined and can be easily iterated. The vectors $a_k$ having been computed,
Proposition 5.4 provides us with the evolution of the frame $g_k$, which
describes the rotation of the top completely. So, given $(m_0,a_0)$, the 
rotation of the top is determined uniquely. Due to (\ref{dLT rest aux3}) one
can take two consecutive positions $(a_0,a_1)$ of the axis as the initial 
conditions as well. Fig.1 demonstrates a typical discrete time precession of the
axis. Compare this with the classical continuous time pictures in \cite{KS},
\cite{A}.

\begin{figure}[tb]
\begin{center} 
\epsfig{file=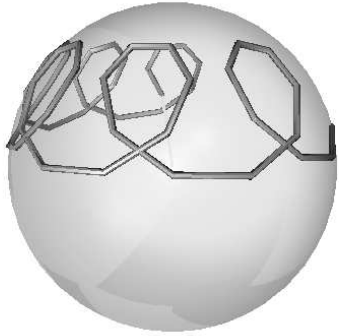,width=60mm} 
\caption{Evolution of the axis of the discrete spinning top} 
\end{center}
\end{figure}

The motion of the discrete time Lagrange top can be viewed using a web--browser.
The Java--applet has been written by Ulrich Heller and can be 
found on the web page
\begin{center}
\begin{verbatim} 
http://www-sfb288.math.tu-berlin.de/~bobenko/bobenko.html
\end{verbatim}
\end{center}
The applet presents an animated spinning top described by the formulas of the 
present paper.

\section{Conclusion}
We took an opportunity of elaborating an integrable discretization of the
Lagrange top to study in a considerable detail the general theory of
discrete time Lagrangian mechanics on Lie groups. We consider this theory
as an important source of symplectic and, more general, Poisson maps.
Moreover, from some points of view the variational (Lagrangian) structure
is even more fundamental and important than the Poisson (Hamiltonian) one
(cf. \cite{HMR}, \cite{MPS}, where a similar viewpoint is represented).

It is somewhat astonishing that this construction is able to produce
{\it integrable} discrete time systems, since integrability is not
built in it {\it a priori}. Nevertheless, we extend the Moser--Veselov's
list \cite{V}, \cite{MV} of integrable discrete time Lagrangian systems with
a new item, namely, an integrable discrete time Lagrange top. It seems that
this list may be further continued.

In finding this new discrete time mechanical system an analogy with
some differential--geometric notions was very instructive. Also these
interrelations between integrable differential geometry and integrable
mechanics, both continuous and discrete, deserve to be studied further.

Let us mention also some more concrete problems connected with this work.
First of all, the discrete time Lax representations found here call for
being understood both from the $r$--matrix point of view \cite{RSTS}, \cite{S} 
and from the point of view of matrix factorizations \cite{MV} (unfortunately, 
these two schemes, being in principle closely related, still could not be 
merged into a unified one). Futher, the discrete 
time dynamics should be integrated in terms of elliptic functions. The
methods of the finite--gap theory will be useful here \cite{RM}. Finally, it 
would be important to elaborate a variational interpretation of different 
integrable discretizations of the Euler top found in \cite{BLS}.

\begin{appendix}
\setcounter{equation}{0}
\section{Notations}
We fix here some notations and definitions used throughout the paper.

Let $G$ be a Lie group with the Lie algebra $\gog$, and let $\gog^*$ be a dual 
vector space to $\gog$. We identify $\gog$ and $\gog^*$ with the tangent space 
and the cotangent space to $G$ in the group unity, respectively:
\[
\gog=T_e G\;,\qquad \gog^*=T^*_e G\;.
\]
The pairing between the cotangent and the tangent spaces $T^*_g G$ and 
$T_g G$ in an arbitrary point $g\in G$ is denoted by 
$\langle\cdot,\cdot\rangle$. The left and right translations in the group
are the maps $L_g\,,R_g\,:G\mapsto G$ defined by
\[
L_g\,h=gh\;,\qquad R_g\,h=hg\qquad\forall h\in G\;,
\]
and $L_{g*}\,$, $R_{g*}$ stand for the differentials of these maps:
\[
L_{g*}\,:\,T_h G\mapsto T_{gh} G\;, \qquad R_{g*}\,:\,T_h G\mapsto T_{hg} G\;.
\]
We denote by
\[
{\rm Ad}\,g = L_{g*}R_{g^{-1}*}\,:\gog\mapsto\gog
\]
the adjoint action of the Lie group $G$ on its Lie algebra $\gog=T_e G$. 
The linear operators
\[
L_g^*\,:\, T^*_{gh} G\mapsto T^*_h G\;, \qquad
R_g^*\,:\, T^*_{hg} G\mapsto T^*_h G
\]
are conjugated to $L_{g*}\,$, $R_{g*}\,$, respectively, via the pairing
$\langle\cdot,\cdot\rangle$:
\[
\langle L_g^*\xi,\eta\rangle=\langle\xi,L_{g*}\eta\rangle\quad{\rm for}
\quad \xi\in T^*_{gh} G\,,\;\; \eta\in T_h G\;,
\]
\[
\langle R_g^*\xi,\eta\rangle=\langle\xi,R_{g*}\eta\rangle\quad{\rm for}
\quad \xi\in T^*_{hg} G\,,\;\; \eta\in T_h G\;.
\]
The coadjoint action of the group
\[
{\rm Ad}^*\,g = L^*_{g}R^*_{g^{-1}}\,:\gog^*\mapsto\gog^*
\]
is conjugated to ${\rm Ad}\,g$ via the pairing $\langle\cdot,\cdot\rangle$:
\[
\langle {\rm Ad}^*\, g\cdot\xi,\eta\rangle=\langle\xi, 
{\rm Ad}\,g\cdot\eta\rangle\quad{\rm for}
\quad \xi\in\gog^*\,,\;\; \eta\in\gog\;.
\]
The differentials of ${\rm Ad}\,g$ and of ${\rm Ad}^*\,g$ with respect to $g$
in the group unity $e$ are the operators 
\[
{\rm ad}\,\eta\,:\gog\mapsto\gog \qquad{\rm and}\qquad
{\rm ad}^*\,\eta\,:\gog^*\mapsto\gog^*\;,
\]
respectively, also conjugated via the pairing $\langle\cdot,\cdot\rangle$:
\[
\langle{\rm ad}^*\,\eta\cdot\xi,\zeta\rangle=\langle\xi,{\rm ad}\,\eta\cdot\zeta
\rangle \qquad \forall \xi\in\gog^*\,,\;\;\zeta\in\gog\;.
\]
The action of ad is given by applying the Lie bracket in $\gog$:
\[
{\rm ad}\,\eta\cdot\zeta=[\eta,\zeta]\;, \quad \forall\zeta\in\gog\;.
\]

Finally, we shall need the notion of gradients of functions on vector spaces
and on manifolds. If $\cX$ is a vector space, and $f:\cX\mapsto{\Bbb R}$ is a 
smooth function, then the gradient $\nabla f:\cX\mapsto\cX^*$ is defined via
the formula
\[
\langle\nabla f(x),y\rangle=\left.\frac{d}{d\epsilon}\,f(x+\epsilon y)
\right|_{\epsilon=0}\;, \qquad \forall y\in\cX\;.
\]
Similarly, for a function $f:G\mapsto{\Bbb R}$ on a smooth manifold $G$ its
gradient $\nabla f\,:\,G\mapsto T^* G$ is defined in the following way: for an
arbitrary $\dot{g}\in T_g G$ let $g(\epsilon)$ be a curve in $G$ through
$g(0)=g$ with the tangent vector $\dot{g}(0)=\dot{g}$. Then
\[
\langle\nabla f(g),\dot{g}\rangle=\left.\frac{d}{d\epsilon}\,f(g(\epsilon))
\right|_{\epsilon=0}\;.
\]
If $G$ is a Lie group, then two convenient ways to define a curve in $G$ through
$g$ with the tangent vector $\dot{g}$ are the following:
\[
g(\epsilon)=e^{\epsilon\eta}g\;,\quad \eta=R_{g^{-1}*}\,\dot{g}\;,
\]
and
\[
g(\epsilon)=ge^{\epsilon\eta}\;,\quad \eta=L_{g^{-1}*}\,\dot{g}\;,
\]
which allows to establish the connection of the gradient $\nabla f$ with
the (somewhat more convenient) notions of the left and the right Lie derivatives
of a function $f:G\mapsto{\Bbb R}$:
\[
\nabla f(g)=R_{g^{-1}}^*\,df(g)=L_{g^{-1}}^*\,d\,'f(g)\;.
\] 
Here $df(g):G\mapsto\gog^*$ and $d\,'f(g):G\mapsto\gog^*$ are defined via 
the formulas
\[
\langle df(g),\eta\rangle=\left.\frac{d}{d\epsilon}\,f(e^{\epsilon\eta}g)
\right|_{\epsilon=0}\;,\qquad \forall \eta\in\gog\;,
\]
\[
\langle d\,'f(g),\eta\rangle=\left.\frac{d}{d\epsilon}\,f(ge^{\epsilon\eta})
\right|_{\epsilon=0}\;,\qquad \forall \eta\in\gog\;.
\]

\setcounter{equation}{0}
\section{Lagrangian equations of motion}

\noindent {\small
\begin{tabular}{|l|l|}\hline
 & \\
\multicolumn{1}{|c|}{CONTINUOUS TIME} & \multicolumn{1}{|c|}{DISCRETE TIME}\\ 
 & \\ \hline\hline   
\multicolumn{2}{|c|}{}\\ 
\multicolumn{2}{|c|}{General Lagrangian systems}\\ 
\multicolumn{2}{|c|}{} \\ \hline
 & \\
$\bL(g,\dot{g})$ & $\bbL(g_k,g_{k+1})$ \\ 
 & \\
$\left\{\begin{array}{l} \Pi=\nabla_{\dot{g}}\bL\\
                                           \dot{\Pi}=\nabla_g\bL
                          \end{array}\right.$ &
                $\left\{\begin{array}{l} \Pi_k=-\nabla_1\bbL(g_k,g_{k+1})\\
                                        \Pi_{k+1}=\nabla_2\bbL(g_k,g_{k+1})
                          \end{array}\right.$ \\
 & \\
\hline\hline \multicolumn{2}{|c|}{} \\
\multicolumn{2}{|c|}{Left trivialization: $\;\;M=L_g^*\Pi$} \\
\multicolumn{2}{|c|}{} \\ \hline
 & \\ 
$\bL(g,\dot{g})=\bL^{(l)}(g,\Omega)$ & 
$\bbL(g_k,g_{k+1})=\bbL^{(l)}(g_k,W_k)$ \\ 
$\Omega=L_{g^{-1}*}\dot{g}$ & $W_k=g_k^{-1}g_{k+1}$  \\ 
 & \\
$M=L_g^*\Pi=\nabla_{\Omega}\bL^{(l)}$ & 
$M_k=L_{g_k}^*\Pi_k=d\,'_{\!W}\bbL^{(l)}(g_{k-1},W_{k-1})$ \\ 
 & \\
$\left\{\begin{array}{l}
\dot{M}={\rm ad}^*\,\Omega\cdot M+d\,'_{\!g}\bL^{(l)}\\ 
\dot{g}=L_{g*}\Omega\end{array}\right.$ &
$\left\{\begin{array}{l}
{\rm Ad}^*\,W_k^{-1}\cdot M_{k+1}=M_k+d\,'_{\!g}\bbL^{(l)}(g_k,W_k)\\ 
g_{k+1}=g_kW_k\end{array}\right.$ \\ 
 & \\
\hline \multicolumn{2}{|c|}{} \\
\multicolumn{2}{|c|}{Left trivialization, left symmetry reduction: 
$\;\;M=L_g^*\Pi\,,\;\; P={\rm Ad}\,g^{-1}\cdot\zeta$} \\
\multicolumn{2}{|c|}{} \\ \hline
 & \\
$\bL(g,\dot{g})=\cL^{(l)}(P,\Omega)$ &
$\bbL(g_k,g_{k+1})=\Lambda^{(l)}(P_k,W_k)$ \\ 
$\Omega=L_{g^{-1}*}\dot{g}\,,\;\;P={\rm Ad}\,g^{-1}\cdot\zeta$ & 
$W_k=g_k^{-1}g_{k+1}\,,\;\;P_k={\rm Ad}\,g_k^{-1}\cdot\zeta$\\ 
 & \\
$M=L_g^*\Pi=\nabla_{\Omega}\cL^{(l)}$ & 
$M_k=L_{g_k}^*\Pi_k=d\,'_{\!W}\Lambda^{(l)}(P_{k-1},W_{k-1})$ \\ 
 & \\
$\left\{\begin{array}{l}
\dot{M}={\rm ad}^*\,\Omega\cdot M+{\rm ad}^*\,P\cdot\nabla_P\cL^{(l)}\\ 
\dot{P}=[P,\Omega]\end{array}\right.$ &
 $\left\{\begin{array}{l}
{\rm Ad}^*\,W_k^{-1}\cdot M_{k+1}=M_k+{\rm ad}^*P_k\cdot
\nabla_P\Lambda^{(l)}(P_k,W_k)\\ 
P_{k+1}={\rm Ad}\,W_k^{-1}\cdot P_k\end{array}\right.$ \\
 & \\
\hline\hline 
\end{tabular}

\noindent
\begin{tabular}{|l|l|}\hline
\multicolumn{2}{|c|}{} \\
\multicolumn{2}{|c|}{Right trivialization: $\;\;m=R_g^*\Pi$} \\
\multicolumn{2}{|c|}{} \\ \hline
 & \\ 
$\bL(g,\dot{g})=\bL^{(r)}(g,\omega)$ & 
$\bbL(g_k,g_{k+1})=\bbL^{(r)}(g_k,w_k)$ \\ 
$\omega=R_{g^{-1}*}\dot{g}$ & $w_k=g_{k+1}g_k^{-1}$  \\ 
 & \\
$m=R_g^*\Pi=\nabla_{\omega}\bL^{(r)}$ & 
$m_k=R_{g_k}^*\Pi_k=d_{w}\bbL^{(r)}(g_{k-1},w_{k-1})$ \\ 
 & \\
$\left\{\begin{array}{l}
\dot{m}=-{\rm ad}^*\,\omega\cdot m+d_g\bL^{(r)}\\ 
\dot{g}=R_{g*}\omega\end{array}\right.$ &
$\left\{\begin{array}{l}
{\rm Ad}^*\,w_k\cdot m_{k+1}=m_k+d_g\bbL^{(r)}(g_k,w_k)\\ 
g_{k+1}=w_kg_k\end{array}\right.$ \\ 
 & \\
\hline  \multicolumn{2}{|c|}{} \\
\multicolumn{2}{|c|}{Right trivialization, right symmetry reduction:
$\;\;m=R_g^*\Pi\,,\;\; a={\rm Ad}\,g\cdot\zeta$} \\ 
\multicolumn{2}{|c|}{} \\ \hline
 & \\
$\bL(g,\dot{g})=\cL^{(r)}(a,\omega)$ &
$\bbL(g_k,g_{k+1})=\Lambda^{(r)}(a_k,w_k)$ \\ 
$\omega=R_{g^{-1}*}\dot{g}\,,\;\;a={\rm Ad}\,g\cdot\zeta$ & 
$w_k=g_{k+1}g_k^{-1}\,,\;\;a_k={\rm Ad}\,g_k\cdot\zeta$\\ 
 & \\
$m=R_g^*\Pi=\nabla_{\omega}\cL^{(r)}$ & 
$m_k=R_{g_k}^*\Pi_k=d_w\Lambda^{(r)}(a_{k-1},w_{k-1})$ \\ 
 & \\
$\left\{\begin{array}{l}
\dot{m}=-{\rm ad}^*\,\omega\cdot m-{\rm ad}^*\,a\cdot\nabla_a\cL^{(r)}\\ 
\dot{a}=[\omega,a]\end{array}\right.$ &
 $\left\{\begin{array}{l}
{\rm Ad}^*\,w_k\cdot m_{k+1}=m_k-{\rm ad}^*a_k\cdot
\nabla_a\Lambda^{(r)}(a_k,w_k)\\ 
a_{k+1}={\rm Ad}\,w_k\cdot a_k\end{array}\right.$ \\
 & \\
\hline\hline
\end{tabular}
}
\vspace{5mm}

The relation between the continuous time and the discrete time equations 
is established, if we set
\[
g_k=g\;,\qquad g_{k+1}=g+\varepsilon\dot{g}+O(\varepsilon^2)\;,\qquad
\bbL(g_k,g_{k+1})=\varepsilon\bL(g,\dot{g})+O(\varepsilon^2)\;;
\]
\[
P_k=P\;,\qquad W_k=\ed+\varepsilon\Omega+O(\varepsilon^2)\;,\qquad
\Lambda^{(l)}(P_k,W_k)=\varepsilon\cL^{(l)}(P,\Omega)+O(\varepsilon^2)\;;
\]
\[
a_k=a\;,\qquad w_k=\ed+\varepsilon\omega+O(\varepsilon^2)\;,\qquad
\Lambda^{(r)}(a_k,w_k)=\varepsilon\cL^{(r)}(a,\omega)+O(\varepsilon^2)\;.
\]

\setcounter{equation}{0}
\section{On $SU(2)$ and $su(2)$}

The Lie group $G=SU(2)$ consists of complex $2\times 2$
matrices $g$ satisfying the condition $gg^*=g^*g=\ed$, where $\ed$ is the 
group unit, i.e. the $2\times 2$ unit matrix, and $^*$ denotes the
Hermitean conjugation, i.e. $g^*=\bar{g}^T$. In components:
\begin{equation}\label{quat}
g=\left(\begin{array}{cc} \alpha & \beta \\ -\bar{\beta} & \bar{\alpha}
\end{array}\right)=
\left(\begin{array}{cc} a+ib & c+id \\ -c+id & a-ib \end{array}\right)\;,
\end{equation}
where
\[
\alpha=a+ib\,,\;\beta=c+id\in{\Bbb C}\;,
\]
and
\begin{equation}\label{norm}
|\alpha|^2+|\beta|^2=a^2+b^2+c^2+d^2=1\;.
\end{equation}
The tangent space $T_eSU(2)$ is the Lie algebra $\gog=su(2)$ consisting of 
complex $2\times 2$ matrices $\eta$ such that $\eta+\eta^*=0$. In components,
\begin{equation}\label{im quat}
\eta=\left(\begin{array}{cc} ib & c+id \\ -c+id & -ib \end{array}\right)\;.
\end{equation}
The Lie bracket in $su(2)$ is the usual matrix commutator. 

Let us introduce the following notations: for an arbitrary matrix
$g$ of the form (\ref{quat}), not necessary belonging to $SU(2)$, set
\[
\Re(g)=\left(\begin{array}{cc} a & 0 \\ 0 & a \end{array}\right)\;,\quad
\Im(g)=\left(\begin{array}{cc} ib & c+id \\ -c+id & -ib \end{array}\right)\;,
\]
so that $\Re(g)$ is a scalar real matrix, and $\Im(g)\in su(2)$.

As it is always the case for matrix groups, we have for $g\in SU(2)$, 
$\eta\in su(2)$:
\begin{equation}\label{ops}
L_{g*}\eta=g\eta\;,\quad R_{g*}\eta=\eta g\;,\quad {\rm Ad}\,g\cdot\eta=
g\eta g^{-1}\;.
\end{equation}

If we write (\ref{im quat}) as
\begin{equation}\label{isom}
\eta=\frac{1}{2}\left(\begin{array}{cc} -i\eta_3 & -\eta_2-i\eta_1 \\
 \eta_2-i\eta_1 & i\eta_3 \end{array}\right)\;,
\end{equation}
and put this matrix in a correspondence with the vector
\[
\eta=(\eta_1,\eta_2,\eta_3)^T\in{\Bbb R}^3\;,
\]
then it is easy to verify that this correspondence is an isomorphism
between $su(2)$ and the Lie algebra $\Big({\Bbb R}^3,\,\times\Big)$, where 
$\times$ stands for the vector product. This allows not to distinguish between 
vectors from ${\Bbb R}^3$ and matrices from $su(2)$. In other words, we use
the following basis of the linear space $su(2)$:
\[
\be_1=\frac{1}{2}\left(\begin{array}{cc} 0 & -i \\ -i & 0 \end{array}\right)=
\frac{1}{2i}\,\sigma_1\;, \qquad
\be_2=\frac{1}{2}\left(\begin{array}{cc} 0 & -1 \\ 1 & 0 \end{array}\right)=
\frac{1}{2i}\,\sigma_2\;,
\]
\begin{equation}\label{e's}
\be_3=\frac{1}{2}\left(\begin{array}{cc} -i & 0 \\ 0 & i \end{array}\right)=
\frac{1}{2i}\,\sigma_3\;,
\end{equation}
where $\sigma_j$ are the Pauli matrices.

We supply $su(2)$ with 
the scalar product $\langle\cdot,\cdot\rangle$ induced from ${\Bbb R}^3$. 
It is easy to see that in the matrix form it may be represented as
\begin{equation}\label{scal pr in alg}
\langle \eta,\zeta\rangle = -2\,{\rm tr}(\eta\zeta)=2\,{\rm tr}(\eta\zeta^*)\;.
\end{equation}
This scalar product allows us to identify the dual space $su(2)^*$ with $su(2)$
itself, so that the coadjoint action of the algebra becomes the usual Lie
bracket with minus:
\begin{equation}\label{coad}
{\rm ad}^*\,\eta\cdot\xi=[\xi,\eta]=-{\rm ad}\,\eta\cdot\xi\;.
\end{equation}
We use a formula similar to (\ref{scal pr in alg}) to define a scalar product 
of two arbitrary complex $2\times 2$ matrices:
\begin{equation}\label{scal pr gen}
\langle g_1,g_2\rangle =2\,{\rm tr}(g_1g_2^*)\;.
\end{equation}
(In particular, the square of the norm of every matrix $g\in SU(2)$ is equal 
to 4). The 
formula (\ref{scal pr gen}) gives us a left-- and right--invariant scalar 
product in all tangent spaces $T_gG$. Indeed, to see, for instance, the left
invariance, let $g\eta,g\zeta\in T_gSU(2)$ (here $g\in SU(2)$, $\eta,\zeta\in
su(2)$). Then
\[
\langle g\eta,g\zeta\rangle=2\,{\rm tr}(g\eta\zeta^*g^*)=
2\,{\rm tr}(\eta\zeta^*)=\langle\eta,\zeta\rangle\;.
\]
This scalar product allows to identify the cotangent spaces $T^*_gG$ with the 
tangent spaces $T_gG$. It follows easily that:
\begin{equation}\label{ops*}
L_g^*\xi=g^{-1}\xi\;, \quad R_g^*\xi=\xi g^{-1}\;, \quad
{\rm Ad}^*\,g\cdot\xi=g^{-1}\xi g\;.
\end{equation}
(in these formulas $g\in SU(2)$, so that $g^{-1}=g^*$).

Let us now formulate several simple properties of $SU(2)$ and $su(2)$ which
will be used later on.
\begin{lemma}
For an arbitrary $g\in SU(2)$: 
\begin{equation}\label{group el as rot}
g=\cos(\theta)\,\ed+\sin(\theta)\,\zeta\qquad{\rm with}
\qquad\zeta\in su(2)\;,\quad \langle\zeta,\zeta\rangle=4\;.
\end{equation}
The adjoint action of $SU(2)$ on $su(2)$ has in these notations the following 
geometrical interpretation:
$g\eta g^{-1}$ is a rotation of the vector $\eta$ around the vector $\zeta$ 
by the angle $2\theta$.
\end{lemma} 

This interpretation makes $SU(2)$ very convenient for describing rotations in 
${\Bbb R}^3$ (in some respects more convenient than the standard use of $SO(3)$ 
in this context). Since by rotations only the vectors on the rotation axis 
remain fixed, we see that for the case $G=SU(2)$
\[
G^{[\zeta]}=G^{(\zeta)}\;.
\]

In a different way, the previous lemma may be formulated as follows.
\begin{lemma}
For $g\in SU(2)$, if
\begin{equation}\label{Im part}
2\,\frac{\Im(g)}{{\rm tr}(g)}=\xi\;,
\end{equation}
then
\begin{equation}\label{reconstruct}
g=\frac{{\rm tr}(g)}{2}\,(\ed+\xi)\;,\qquad
{\rm tr}(g)=\frac{2}{\sqrt{1+\langle\xi,\xi\rangle/4}}\;.
\end{equation}
\end{lemma}

We have also the following simple connection between the matrix multiplication
and the commutator in $su(2)$ .
\begin{lemma}
For $\eta,\zeta\in su(2)$ their matrix product has the form {\rm(\ref{quat})}, 
and
\begin{equation}\label{im quat prod}
\eta\zeta=-\frac{1}{4}\,\langle \eta,\zeta\rangle\ed+
\frac{1}{2}\,[\eta,\zeta]\;.
\end{equation}
\end{lemma}

In particular, the following corollary is important:
\begin{equation}\label{trick}
\langle \eta,\zeta\rangle=0\;\Rightarrow\;\eta\zeta=-\zeta\eta\;.
\end{equation}

\end{appendix}



\begin{thebibliography}{WWW}
\bibitem[AL]{AL}  M.Ablowitz, J.Ladik.  A nonlinear difference scheme and
 inverse scattering. {\em Stud. Appl. Math.}  {\bf 55} (1976) 213--229;
 On solution of a class of nonlinear partial difference equations. 
 {\em Stud. Appl. Math.} {\bf 57} (1977) 1--12.

\bibitem[AM]{AM} M.Adler, P.van Moerbeke. Completely integrable systems,
Euclidean Lie algebras and curves. {\em Adv. Math.} {\bf 38} (1980) 267--317.

\bibitem[A]{A} V.I.Arnold. {\em Mathematical methods of classical mechanics}.
Springer, 1978.

\bibitem[Au]{Au} M.Audin. {\em Spinning tops}. Cambridge University Press, 1996.

\bibitem[B]{B} A.I.Bobenko. Discrete integrable systems and geometry.
{\em Proc. Int. Congress Math. Phys. '97} (to appear).

\bibitem[BLS]{BLS} A.I.Bobenko, B.Lorbeer, Yu.B.Suris. Integrable 
discretizations of the Euler top. {\em J. Math. Phys.} (1998, to appear).

\bibitem[BP]{BP} A.I.Bobenko, U.Pinkall. Discretization of surfaces and
integrable systems. -- In: {\em Discrete integrable geometry and physics}, Eds.  
A.Bobenko and R.Seiler, Oxford University Press, 1998. 

\bibitem[CB]{CB} R.H.Cushman, L.M.Bates. {\em Global aspects of classical
integrable systems}. Birk\-h\"auser, 1997.

\bibitem[DJM]{DJM} F.Date, M.Jimbo,  and T.Miwa. Method for generating 
 discrete soliton equations. I--IV. {\em J. Phys. Soc. Japan} {\bf 51} (1982)
 4116--4124, 4125--4131; {\bf 52} (1983) 761--765, 766--771.

\bibitem[DS]{DS} A.Doliwa, P.Santini. Geometry of discrete curves and
lattices and integrable difference equations. -- In: {\em Discrete integrable 
geometry and physics}, Eds. A.Bobenko and R.Seiler, Oxford University Press, 
1998. 

\bibitem[FT]{FT} L.D.Faddeev, L.A.Takhtadjan. {\em Hamiltonian methods in
the theory of solitons}, Springer, 1987.

\bibitem[G]{G} V.V.Golubev. {\em Lectures on the integration of the equations of
motion of a rigid body about a fixed point}. State Publishing, Moscow, 1953. 

\bibitem[Ha]{Ha} H.Hasimoto. Motion of a vortex filament and its relation
to elastica. {\em J. Phys. Soc. Japan} {\bf 31} (1971) 293--294;\\
A soliton of a vortex filament. {\em J. Fluid Mech.} {\bf 51} (1972) 477--485.

\bibitem[H]{H} R.Hirota. Nonlinear partial difference equations. I--V. 
 {\em J. Phys. Soc. Japan} {\bf 43} (1977)  1423--1433, 2074--2078, 2079--2086;
 {\bf 45} (1978) 321--332; {\bf 46} (1978) 312--319.

\bibitem[HMR]{HMR} D.D.Holm, J.E.Marsden, T.Ratiu. The Euler--Poincare
equations and semidirect products with applications to continuum theories.
{\em Adv. Math.} (1998, to appear).

\bibitem[KS]{KS} F.Klein, A.Sommerfeld. {\em \"Uber die Theorie des Kreisels}.
Teubner, 1965. (Reprint of the 1897--1910 edition).

\bibitem[L]{L} A.E.H.Love. {\em A treatize of the mathematical theory of 
elasticity}. Cambridge, 1892.

\bibitem[LS]{LS} J.Langer, D.Singer. Lagrangian aspects of the Kirchhoff
elastic rod. {\em SIAM Rev.} {\bf 38} (1996) 605--618.

\bibitem[MPS]{MPS} J.E.Marsden, G.W.Patrick, S.Shkoller. Multisymplectic
geometry, variational integrators, and nonlinear PDEs. {\em Commun. Math.
Phys.} (1998, to appear).

\bibitem[MR]{MR} J.E.Marsden, T.S.Ratiu. {\em Introduction to mechanics 
and symmetry}. Springer, 1994.

\bibitem[MRW]{MRW} J.E.Marsden, T.S.Ratiu, A.Weinstein. Semi--direct
products and reduction in mechanics. {\em Trans. Am. Math. Soc.} {\bf 281}
(1984) 147--177;
Reduction and Hamiltonian structures on duals of semidirect products
Lie algebras. {\em Contemp. Math.} {\bf 28} (1984) 55--100.

\bibitem[MS]{MS} J.E.Marsden, J.Scheurle. Lagrangian reduction and the
double spherical pendulum. {\em ZAMP} {\bf 44} (1993) 17--43; \\
The reduced Euler--Lagrange equations. {\em Fields Inst. Comm.} {\bf 1}
(1993) 139--164.

\bibitem[MV]{MV} J.Moser, A.P.Veselov.  Discrete versions of some 
classical integrable systems and factorization of matrix polynomials. 
{\em Commun. Math. Phys.} {\bf 139} (1991) 217--243.

\bibitem[QNCV]{QNCV}  G.Quispel, F.Nijhoff, H.Capel,  and J.Van der Linden 
 Linear integral equations and nonlinear differential--difference equations. 
 {\em Physica A}   {\bf 125} (1984) 344--380.

\bibitem[RM]{RM} T.Ratiu, P.van Moerbeke. The Lagrange rigid body motion.
{\em Ann. Inst. Fourier} {\bf 32} (1982) 211--234.

\bibitem[R]{R} A.G.Reyman. Integrable Hamiltonian systems connected with
graded Lie algebras. {\em J. Sov. Math.} {\bf 19} (1982) 1507--1545.

\bibitem[RSTS]{RSTS} A.G.Reyman,  M.A.Semenov-Tian-Shansky.  Group theoretical 
methods in the theory of finite dimensional integrable systems.
In: {\em Encyclopaedia of mathematical science}, v.16: {\em Dynamical Systems 
VII}, Springer, 1994, 116--225.

\bibitem[Skl]{Skl} E.K.Sklyanin. On some algebraic structures related to the
Yang--Baxter equation. {\em Fuct. Anal. and Appl.} {\bf 16} (1982) 27--34.
 
\bibitem[S]{S} Yu.B.Suris. $R$--matrices and integrable discretizations. --
In: {\em Discrete integrable geometry and physics}, Eds.  A.Bobenko and R.Seiler,
Oxford University Press, 1988.

\bibitem[V]{V} A.P.Veselov. Integrable systems with discrete time and 
difference operators. {\em Funct. Anal. Appl.} {\bf 22} (1988) 1--13.

\bibitem[WM]{WM} J.M.Wendlandt, J.E.Marsden. Mechanical integrators derived
from a discrete variational principle. {\em Physica D} {\bf 106} (1997)
223--246.



\end{thebibliography}
\end{document}